\documentclass[twocolumn,times,pre,amsmath,amssymb,floatfix]{revtex4} 
\usepackage{graphicx}
\usepackage{epsfig}

\begin{document}
\title{Comparing the Usage of Global and Local Wikipedias \\ with Focus on Swedish Wikipedia}
\author{Berit Schreck}
\author{Mirko K\"ampf}
\author{Jan W. Kantelhardt}
\email[]{jan.kantelhardt@physik.uni-halle.de}
\affiliation{Institut f\"ur Physik, Martin-Luther-Universit\"at Halle-Wittenberg,
06099 Halle (Saale), Germany}
\author{Holger Motzkau}
\email[]{holger.motzkau@wikimedia.se}
\affiliation{Wikimedia Sverige, Box 500, SE-101 29 Stockholm, Sweden}
%\author{Lev Muchnik}
%\email[]{levmuchnik@gmail.com}
%\affiliation{Leonard N. Stern School of Business, New York University, New York, USA}

\date{August 8, 2013}

\begin{abstract}
This report summarizes the results of a short-term student research project focused on 
the usage of Swedish Wikipedia.  It is trying to answer the following question:  To what 
extent (and why) do people from non-English language communities use the English Wikipedia 
instead of the one in their local language?  Article access time series and article edit 
time series from major Wikipedias including Swedish Wikipedia are analyzed with various 
tools.
\end{abstract}
\maketitle

\section{Introduction}
Using Wikipedia can refer to accessing (reading) articles, editing articles, participating 
in discussions (which corresponds to reading and editing discussion pages), etc.  
These activities can be done 
by users inside the particular country under consideration, but also from people living outside 
the country and somehow associated with it (by nationality, interest, etc.).  Here we want to
study the relations between English Wikipedia and Swedish Wikipedia.  To get a basis of reference 
and identify special properties of the Swedish Wikipedia and its users, we also study data from
five additional local Wikipedias.  A particular focus is placed on choosing language communities 
in different time zones, so that information can be gained from comparing daily and weekly 
cycles of activity.

We started with a literature review which included online material.  The focus was on 
publications with content related to our main research question.  The itemized literature 
overview can be found in Section \ref{sec:Literature review}.  In particular, the Wikimedia 
Report Card has to be mentioned as a useful statistical tool.  Although there are other 
projects with topics related with parts of our project, we could not extract many useful 
results for our study from their reports.

The outline of the report is as follows.  Section II describes the data base and the lists of
articles (Wikipedia pages) used for the study.  Section III studies correlations between total 
access volumes 
per page and total edit volumes per page for Swedish and English articles.  Section IV presents
diurnal variations of access volume for Swedish and English articles, while more language 
versions are compared in Section V and edit volume variations are also considered in Section VI.
Section VII looks at synchronous edits in Swedish and English articles, and Section VIII 
characterizes growth patterns of Wikipedia content and quality in different languages.  The 
literature review can be found in Section IX.

\section{Wikipedia Data and Article Selection}

The data used in this work are based on publicly available files collected by Domas Mituzas 
during a time frame from October 2008 till the February 2010 [D. Mituzas, wikidata/wikistats,
http://dammit.it/wikistats (2010)].  The collection was processed by Lev Muchnik (Leonard N. 
Stern School of Business, New York University, New York, USA), focusing on the period from
January 1, 2009 to October 21, 2009.  The total number of article access events during time 
frames of one hour were counted, yielding access-rate time series $n^{j}(t), t=0,\ldots ,7055$
for each article $j$.  An access is defined as the loading of an article page.  We cannot 
distinguish between opening an article in a web-browser or downloading of the page by a web 
crawler. 
The edit time series contain time stamps for each edit event for each article.  We also used 
the MediaWiki API, from which we got the edit time series for the whole period the article 
existed.

\begin{figure}[t]
\begin{center}\includegraphics[width=\hsize]{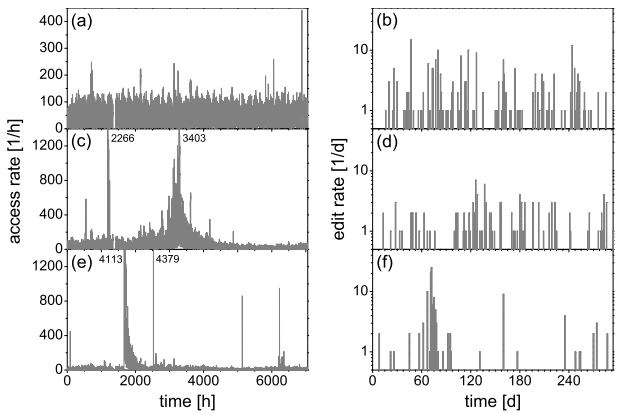}\end{center}
\caption{Examples of Wikipedia (a,c,e) access-rate and (b,d,f) edit-rate time 
series for three selected articles
with (a,b) rather stationary access rates (topic 'Illuminati (book)'), (c,d)
an apparently endogenous burst of activity (peak on May 7, 2009, topic
'Heidelberg'), and (e,f) an exogenous burst of activity (topic 'Amoklauf
Erfurt' (shooting rampage); it peaks on March 11, 2009, as another shooting rampage 
occurred in Winneden).  The left parts show the complete hourly access-rate time series 
(from January 1, 2009, till October 21, 2009; i.~e. for 7056 hours, width of each bar 
is one hour and it represents the number of accesses during this hour) with numbers in the plot 
giving the height of peaks truncated to show baseline fluctuations.  The gap around 
$t=1200$h is a systematical disruption and was found in all records.  Parts (b,d,f) 
show daily edit-rate data (width of each bars is one day and it represents the number of 
edits during this day) for the same three representative 
articles, which were edited 270, 163, and 157 times in total, respectively, during the 
recording period. (Figure taken from K\"ampf et al, Physica A {\bf 391}, 9101 (2012))}
\label{Fig:1}
\end{figure}

Figures 1(a,c,e) show three representative time series of hourly access-rate data $n^{j}(t)$.  
One can distinguish several types of variations: stationary fluctuations [Fig.~\ref{Fig:1}(a)], 
an apparently 'endogenous' burst [Fig.~\ref{Fig:1}(c)] with 
significant precursory activity, and an 'exogenous' burst [Fig.~\ref{Fig:1}(e)] marking the 
dynamic response to a major event.  In addition to random fluctuations and bursting activity, 
we observe daily and weakly activity patterns in the access rates for most Wikipedia articles.  
Periodic minima correspond to the night-time hours, although their positions depend on the 
time zone of the readers and thus on the language of the articles.  Recurring weekly patterns 
are also present in many articles.

Figures 1(b,d,f) show the edit time series of the three representative articles.  In this case, 
the total number of article edits is displayed for each day.  In the report, we denote the 
number of edits per hour by $\epsilon\,^{j}(t), t=0,\ldots ,7055$.  Note that the edit-rate 
shown in Fig.~\ref{Fig:1}(f) peaks on the same day as the access rate of this article (see 
Fig.~\ref{Fig:1}(e)), but there is no such relation for the other two articles.  

For this study, we selected three groups of Wikipedia articles.  Initially we chose the 
articles manually, but then, in order to get more statistical relevance, we did the procedure 
automatically by a computer program.  

{\bf Group 1} contains only Swedish {\it articles of 
local interest}, which do not have a corresponding English article.  To systematically identify 
such Swedish articles, we first compiled a list of all Swedish articles in our data base (from 
2009), then sorted this list by the total number of accesses for each article during our 
observation period, and finally picked the 750 most-accessed articles {\it without} 
an Interwiki link to an English language version.  We focused on the frequently accessed 
articles, since random fluctuations of normalized access and edit activity are weaker compared
with rarely accessed articles and diurnal variations can be studied more reliably.  For example,
the error bar of $n_j(t)$ is $\sqrt{n_j(t)}$, so that the error bar of the normalized time 
series $n_j(t)/\langle n_j(t) \rangle_t$ decreases with increasing average access rate 
$\langle n_j(t) \rangle_t$.

{\bf Group 2} contains 750 Swedish {\it articles of global interest}, which have an 
Interwiki link to an English language version, together with the corresponding English articles
from the main English Wikipedia.  The English Wikipedia was chosen because it is the largest, 
and our main focus was on users possibly looking at the English version rather than the Swedish 
version.  Again, we focused on the articles most frequently accessed in the Swedish version.  

\begin{figure}[t]
(a) \includegraphics[width=0.9\hsize]{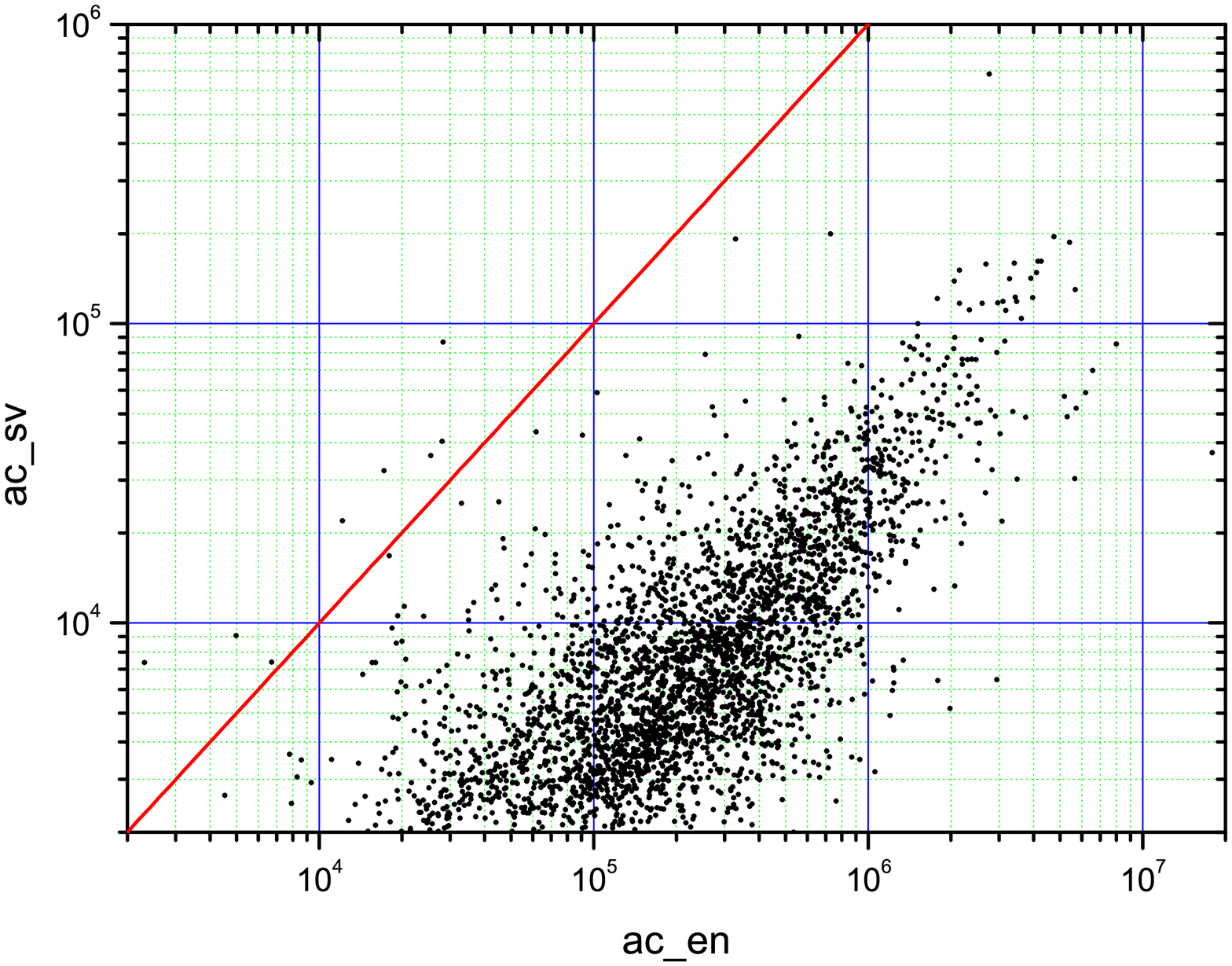}
(b) \includegraphics[width=0.9\hsize]{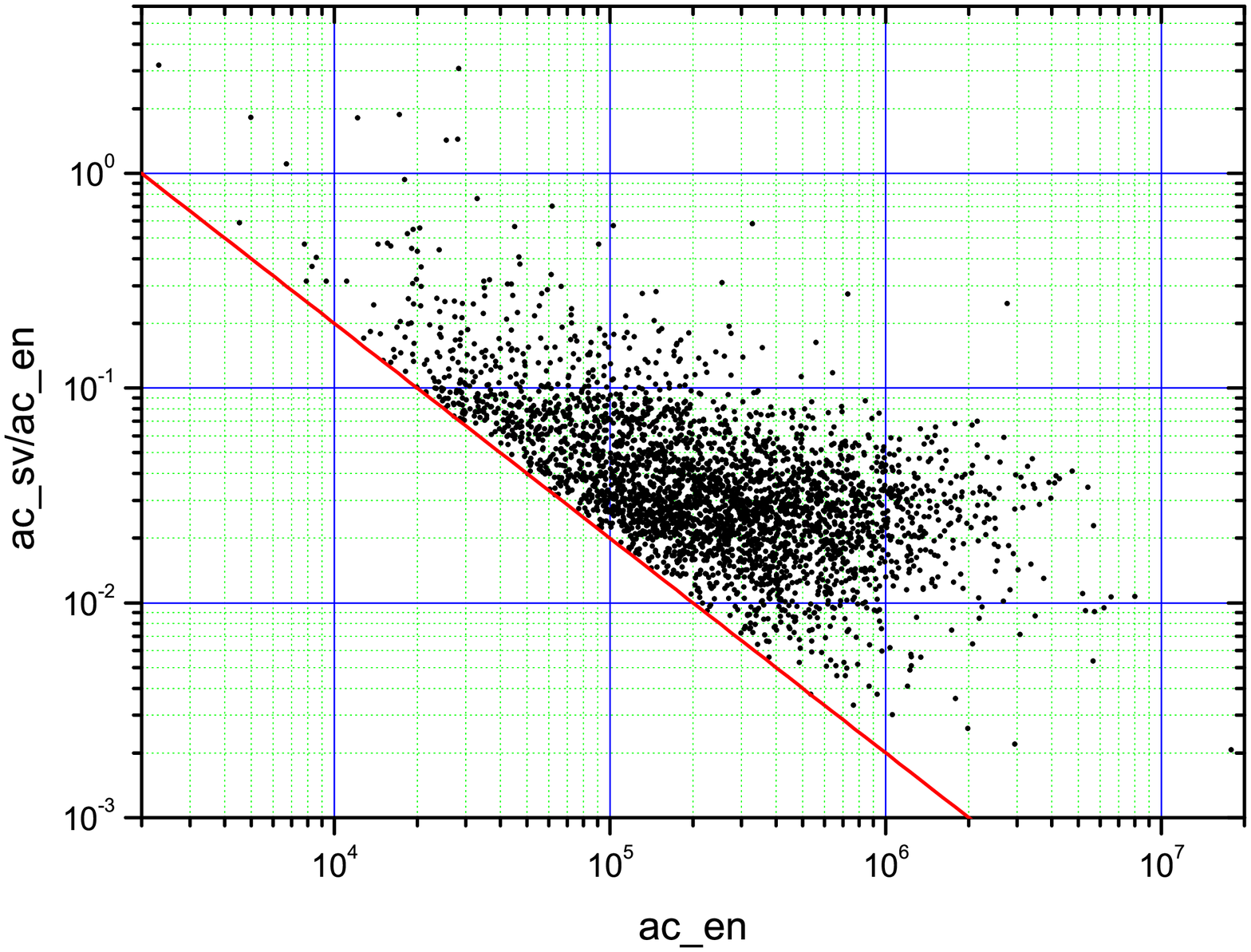}
(c) \includegraphics[width=0.9\hsize]{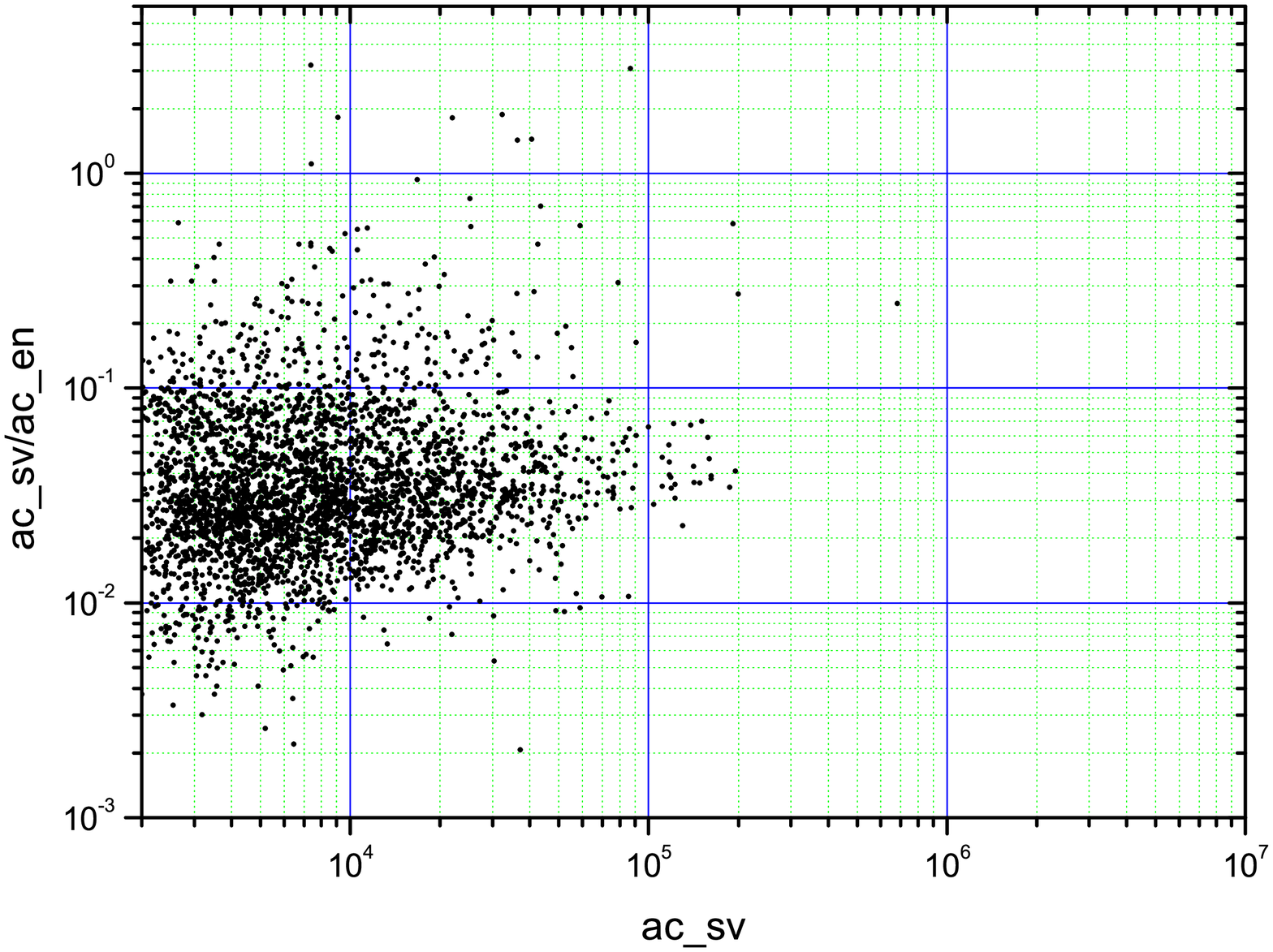}
\caption{(a) Total access volumes of Swedish pages versus total access volumes of corresponding 
English pages. The red line marks equal access volume and is for the orientation.  (b,c) Relative
frequency of access of Swedish version over English version versus (b) access volume in English 
and (c) access volume in Swedish.  The red line in (b) corresponds to 2000 accesses of the Swedish
articles and is our lower cutoff for the inclusion of articles in group 2.}
\end{figure} 

{\bf Group 3} is a subgroup of the second group and defined for international comparison.  
In addition to Swedish and English we chose five other languages and included the articles with 
corresponding versions in all seven Wikipedias.  We chose the Finnish and the Dutch Wikipedia, 
since their total numbers of articles are similar to the Swedish Wikipedia, and they are also
in Europe.  They are thus expected to have similar characteristic features.  The Hebrew 
Wikipedia is also similar to the Swedish regarding the total number of articles, but Israel is 
in a different time zone and switches to daylight-saving time on different days of the year
compared with Europe and the US.  The 
Korean and the Japanese Wikipedias are almost as large (considering the amount of articles) as 
the Swedish one, but located in a time zone with very large difference and no daylight-saving 
time. So we chose them in order to have a reference with different attributes.  For each language 
we created a list of pages with Interwiki links to the original Swedish articles. 

Hourly access rate time series and edit event time series were extracted for each article in each
list.  In order to identify possible differences between normal Wikipedia articles (original 
namespace 0) and other kinds of Wikipedia pages (media pages, discussion pages, etc., namespace 
$\ne 0$), we also compared the results for the full groups 1-3 with results for restricted 
groups 1'-3', in which all pages with namespace $\ne 0$ were excluded.

\begin{figure}[t]
(a) \includegraphics[width=0.9\hsize]{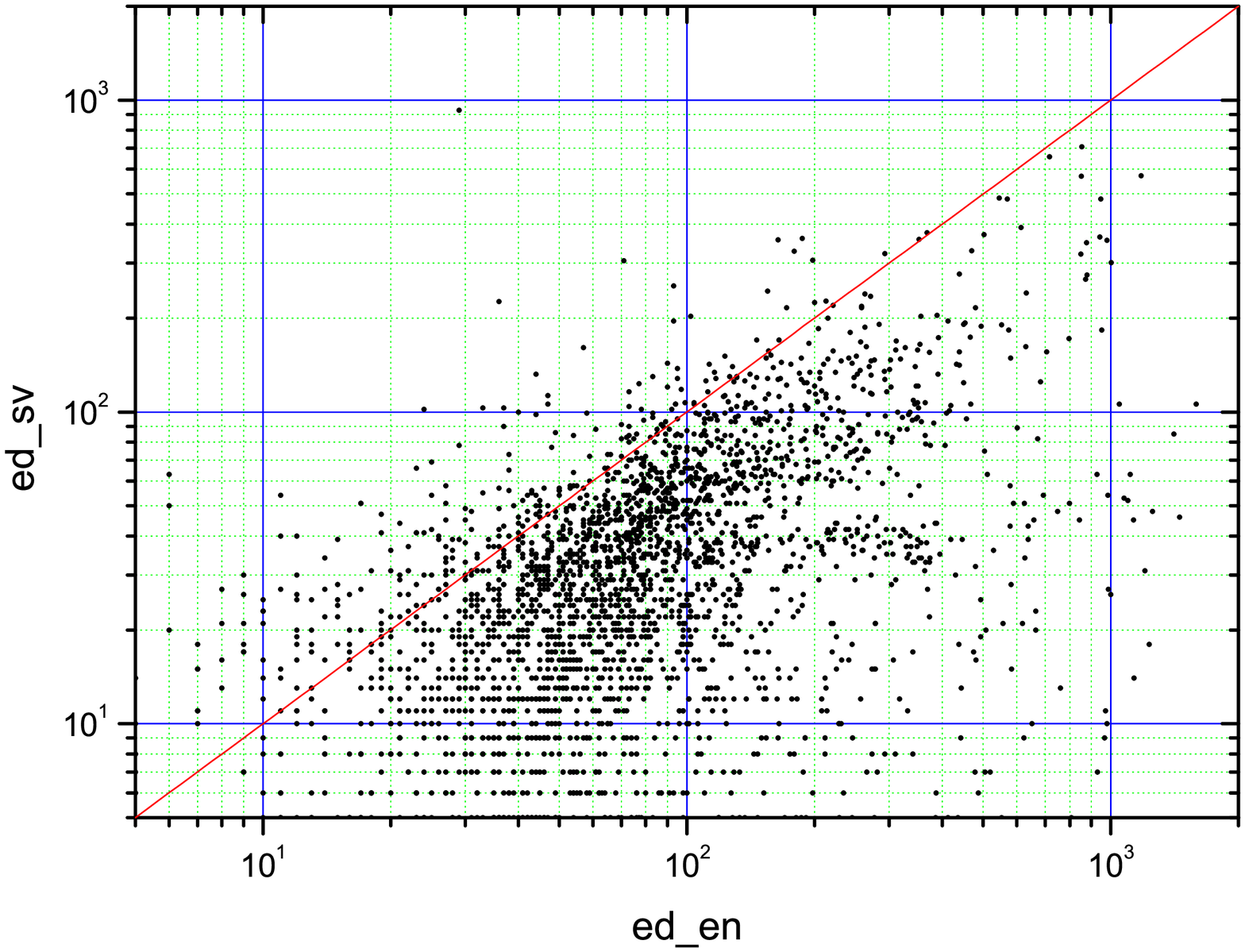}
(b) \includegraphics[width=0.9\hsize]{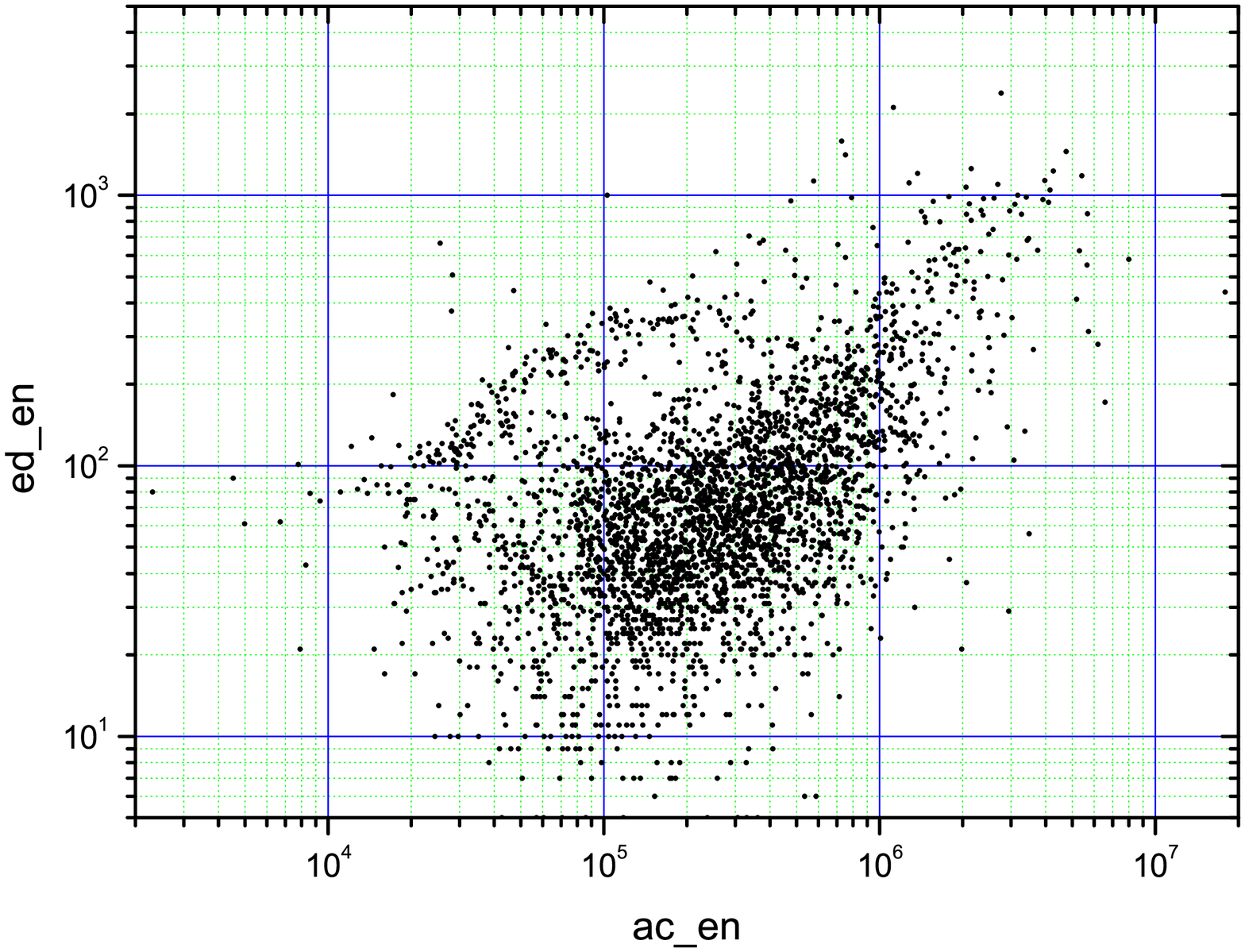}
(c) \includegraphics[width=0.9\hsize]{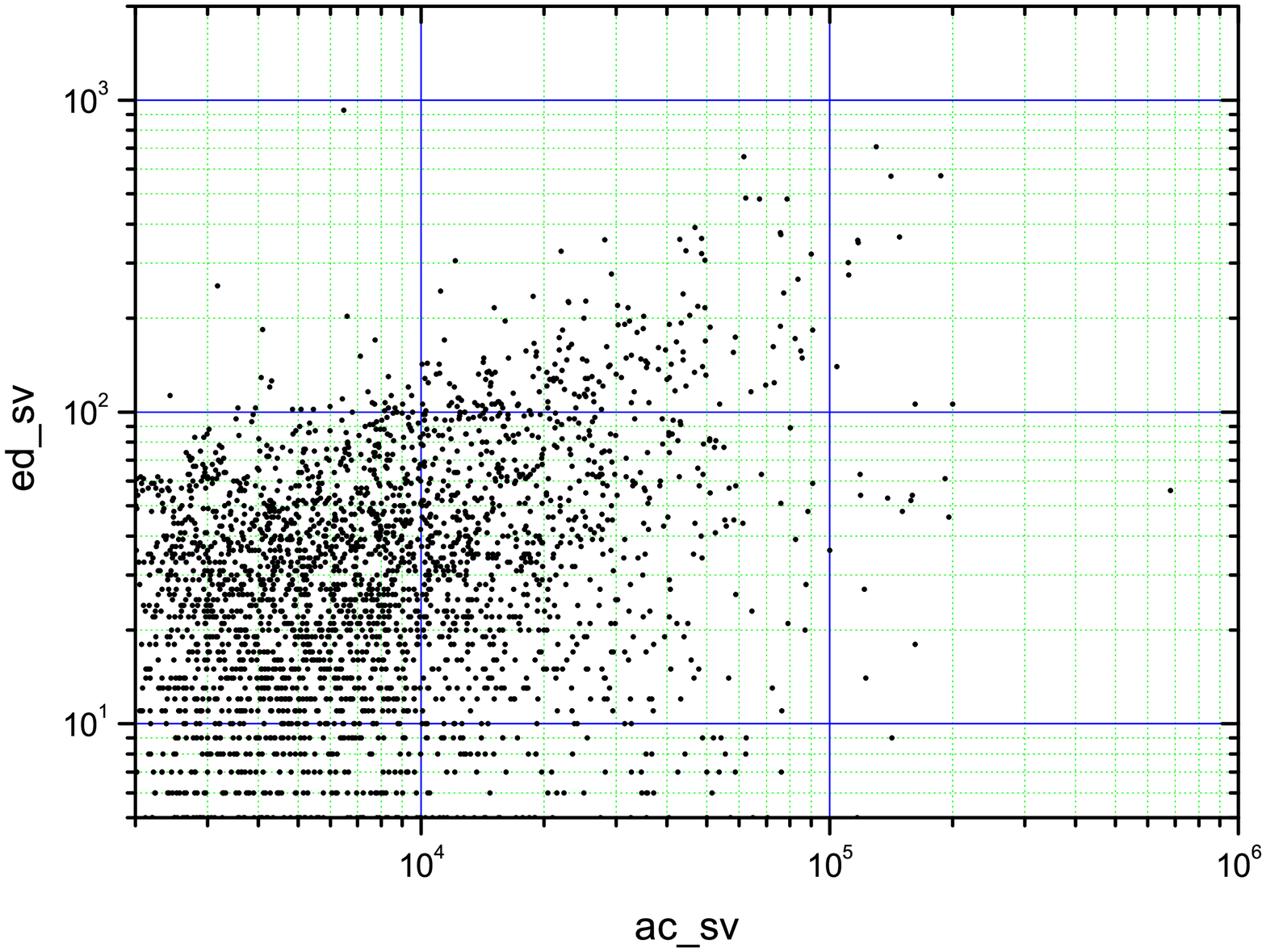}
\caption{(a) Total edit volumes of Swedish pages versus total edit volumes of corresponding 
English pages. The red line marks equal edit volume and is shown for orientation.  (b,c) Edit
versus access volumes for (b) English and (c) Swedish pages.}
\end{figure} 

\section{Correlations of Total Access and Edit Volumes}

In this Section we study total access volume and total edit volume for all articles in group 2,
i.~e., the Swedish articles with corresponding English version, and the corresponding English articles.

Figure 2(a) shows the total number of accesses to the Swedish page versus the total number of 
accesses to the corresponding English page.  In most cases, the English pages are accessed about 
20-50 times more often than the Swedish pages.  Only very few pages were accessed more frequently in 
Swedish than in English (above the red line).  Figures 2(b,c) show the quotient of the total 
number of accesses in Swedish and English Wikipedia versus access volume in English and in Swedish,
respectively.  A clustering can be seen 
in (b), where the quotient is constant around 2-5 percent for the different numbers of 
accesses.  There seems to be another cluster with a larger ratio (around 10 percent) at lower 
English access volumes.  It might represent the pages with general interest but also a particular 
interest in Sweden.  However, the quotient for these pages is still less than one, i.e. there are 
still more accesses on these pages in English. 

Figure 3(a) shows the corresponding results for edit activity of the articles in group 2.  
Again, the English pages are more frequently edited in most cases.  The cross-scatter plots
of edit versus access activity show a sub-cluster of very frequently edited but less frequently
accessed pages in English (Fig. 3(b), top left), which is not so clear in Swedish (Fig. 3(c))).  
This might by partially due to an increased frequency of so-called edit wars in the English 
Wikipedia compared with the Swedish version; possibly edit wars are fought in English more 
often than in Swedish.

\begin{figure}[t]
\centering \includegraphics[width=0.9\hsize]{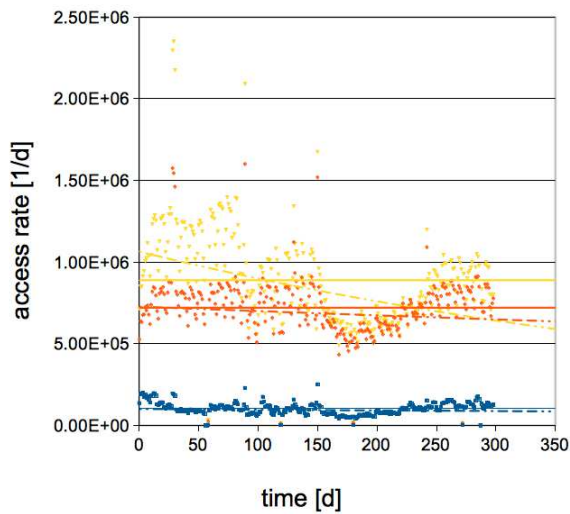}
\caption{Typical trends for changes in access volume of Swedish Wikipedia articles, for which 
an English version is created during the time frame of our study.  Accumulated access volume 
per day is plotted versus time in days since 01.01.2009 as blue, red, and yellow symbols for 
three groups of articles with very large ($N \ge 10^5$, yellow), large ($10^5 > N \ge 10^4$, 
red), and intermediate ($10^4 > N$, blue) total access volume $N$.  The continuous 
lines mark the average values, and the dashed lines are exponential fits.  The drops in access 
frequency between days 150 and 240 are probably due to server problems with counting accesses.  
The strong variations disallow a reliable estimation of growth trends, since the access volume 
seems to be apparently decreasing in the considered time frame.}
\end{figure}

We also tried to study the effect of the creation of an English Wikipedia article on the access 
volume of the corresponding Swedish version which existed already before.  For this approach, 
we identified -- independent of the groups 1-3 described in Section 2 -- all those Swedish 
Wikipedia articles, for which a corresponding (Interwiki-linked) English version was created between 
April 2009 and July 2009.  Although the automated search for such a creation pattern yielded 
several hundreds of articles, many of them had to be dropped in a manual checking procedure
because of pre-existing but differently linked English articles.  Finally, only about one-hundred
articles remained.  The statistics of the access volume changes for the Swedish versions turned out 
to be not significantly different before and after the English version was created.  The main 
reason for this negative results lies in the rather strong variations of article access
volume throughout the observation period.  It was not possible to eliminate the bias of the general
growth of overall Wikipedia access volume (see also Section VIII), because this growth was far 
from constant.  Furthermore, during some month of our observational period significant drops also 
occurred, see Fig. 4.  These drops are partly due to access traffic missing in the server 
statistics.  

Altogether, we can conclude that access volume for typical Swedish articles decreases by less than 
a few percent when an English version is created.  This is a major result, since it shows that 
English articles do not draw away much attention from Swedish articles.  There are further 
implications towards the main question.  Most people read articles in Swedish if they exist, even 
if they might also read the English article.  This raises another question: is the number of 
Wikipedia readers increasing just as there is more information available?  How does this work 
with a usual market?  It contradicts a bit the old economic theories (proven to be incorrect) that 
assumed that people are just consuming more if more products are available.  More detailed research
based on data from longer time periods will be required for answering such questions.  With our 
restricted data base we also could not study the reverse effect, i.e., possible changes of access 
volume of English pages when a Swedish version is created, because statistics are even worse for 
this direction.  We assume that the influence is marginal (if present at all) and could only be 
measured in much larger studies.

\section{Weekly and Diurnal Variations in Groups 1 and 2}

In this Section we compare weekly and diurnal variations of access rates for (i) the Swedish
articles of local interest (group 1), (ii) the Swedish articles on topics of global interest 
(group 2) and (iii) the corresponding English articles (also group 2).  For the weekly variations 
we counted for each day of the week the access in all pages and divided it by the total number of 
accesses of all pages.  For the diurnal variations we did nearly the same, but counted over 
every hour of the days, or, in other words, treated all days of one week the same.

\begin{figure}[t]
\centering \includegraphics[width=0.9\hsize]{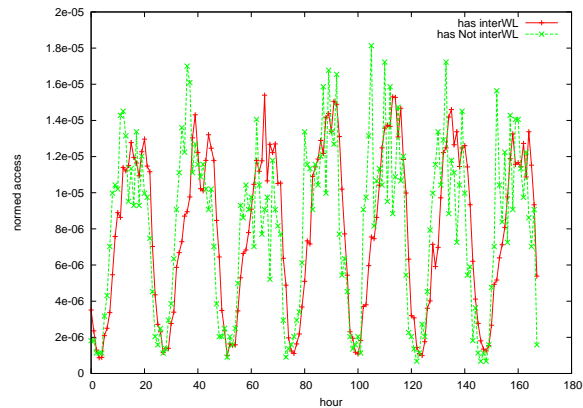}
\caption{Weekly variations for the automatically created lists, (i) group 2 with Swedish pages
of globally interesting topics (with Interwiki links, red) and (ii) group 1 without Interwiki 
links (green).}
\end{figure}

Figure 5 shows the results for the Swedish pages in groups 1 and 2.  There is no significant 
difference in access behaviour for the pages with our without existing English version.  The 
pages with the Interwiki links (global interest) were similarly often accessed during the night
respectively the day as the pages without such links (regional interest).  No distinct group 
of people knowing Swedish (and thus looking at the pages in group 1) but preferring the English 
version (and thus {\it not} looking at the Swedish versions in group 2) can thus be identified 
from studying weekly variations.

\begin{figure}[t]
(a) \includegraphics[width=0.9\hsize]{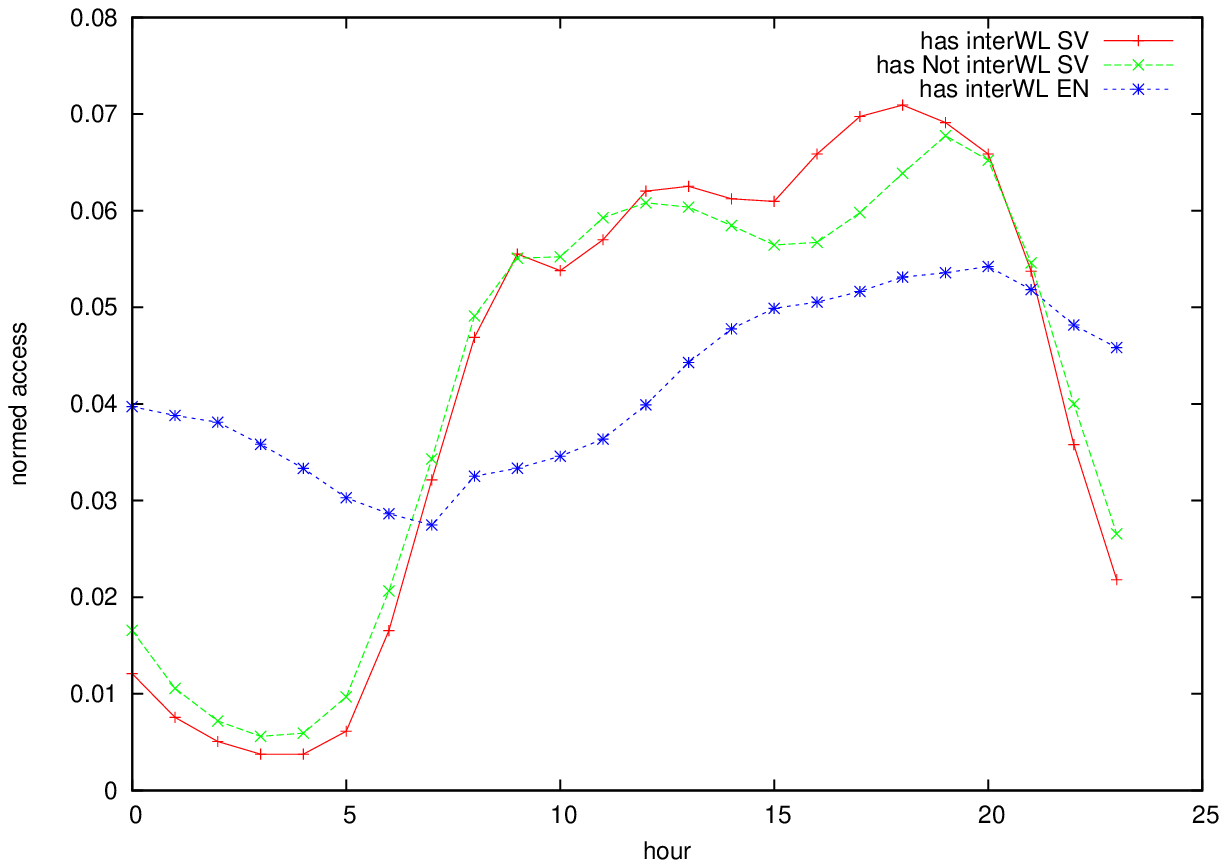}
(b) \includegraphics[width=0.9\hsize]{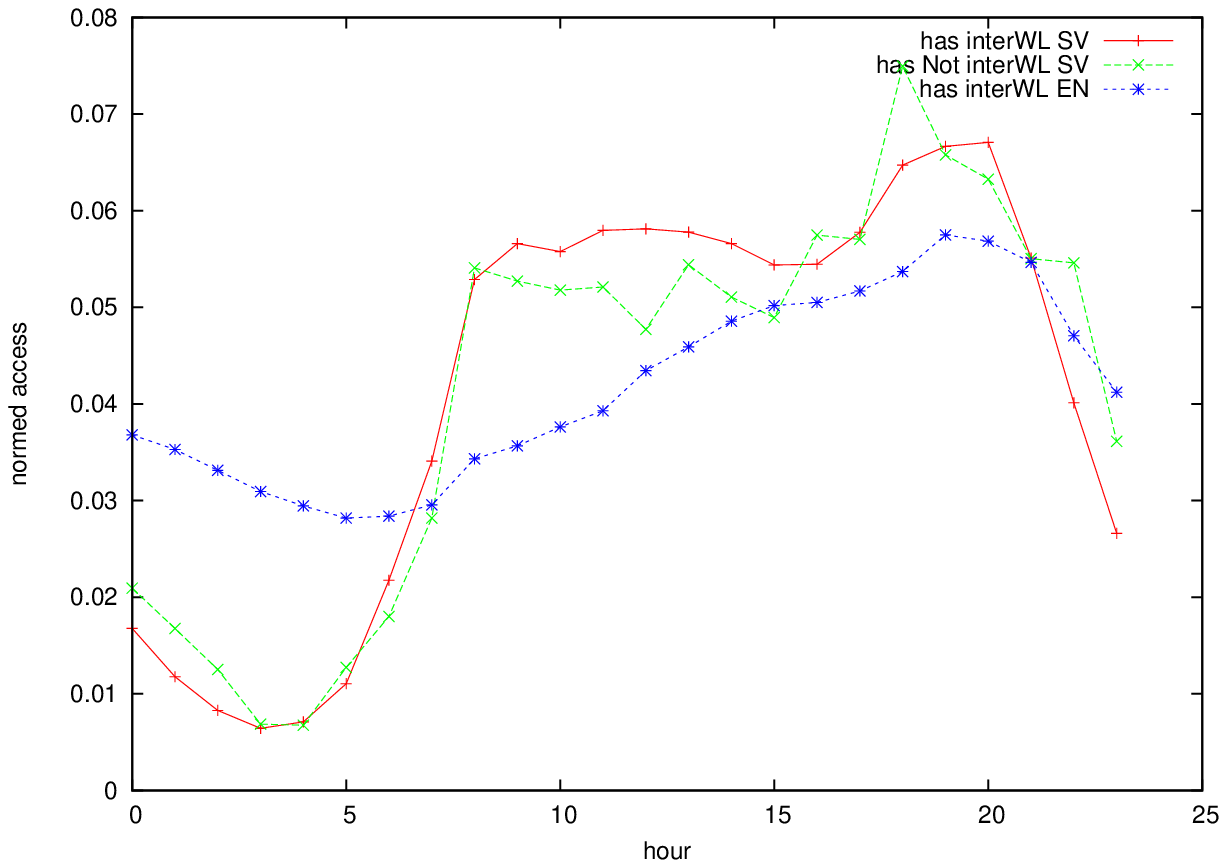}
\caption{Diurnal variations for the lists of Swedish pages with (group 2) as well as without 
(group 1) Interwiki links and the corresponding English pages.  The lists for the curves for (a) 
were created systematically irrespective of the articles' topics, while the lists for (b) were
created by manually choosing appropriate articles of global (group 2) and regional (group 1) 
interest as well as the corresponding English versions of the former.  The standard 
deviations (error bars) for (a) are approximately half as large as those in Fig.~7(c) (i.e., 
between 0.001 and 0.002), since there are 3-4 times more articles in groups 1 and 2 compared 
with group 3.  For (b) the error bars are at least three times larger; therefore most of our 
conclusions are based on (a).}
\end{figure}

Looking at diurnal variations yields better statistics and can thus identify slight differences
between the two groups.  Figure 6 shows that the diurnal variations for the Swedish pages with 
and without links are close but not completely identical.  Diurnal variations for the English 
pages, on the other hand, are very different and much less, since English is used all over
the world.  The first as well as the last hours in the plot represent the night (time is 
according to UTC).  Obviously, 
one can find much less access for the Swedish versions there.  This is logical because during 
the night most of the people are asleep and do not access Wikipedia.  Nevertheless the amount 
of access is never zero.  However, the drop is strongest for the Swedish pages in group 2, 
which have a corresponding English version.  Therefore, if we assumed that a large fraction of 
the people reading Swedish articles at 3 or 4 o'clock at night live outside Europe, we could 
conclude that Swedish versions are more frequently preferred by Swedish users in Europe than 
by Swedish users 
in other parts of the world.  At least approximately one third of the Swedish users in other 
parts of the world prefer to look at the English version, if it exists (the ratio is estimated 
at night-time hours 3 and 4 o'clock in Fig. 6(a)).  For manually selected articles no such 
differences could be established (see Fig. 6(b)), because statistics are worse.

In general, Swedish pages in groups 1 and 2 show a typical diurnal profile.  In the hours of the 
early morning there is a rising of the number of accesses as more and more people get up and 
check something on Wikipedia.  There are two typical maxima:  one around noon time and the other 
one around 7 p.m., corresponding to the work-life rhythm of the people.  So in the Swedish 
Wikipedia the day-night-cycle is very distinctive, but not in the English Wikipedia.  The slight
shift in the position of the second peak for Swedish articles in groups 1 and 2 may be also 
interesting, since it might indicate that people tend to look at articles of more regional 
character later in the evening.

\section{Diurnal Variations of Access Rate in Group 3}

In this Section the diurnal variations of access rates are studied for each of the seven 
languages in group 3. 

\begin{figure}[t]
(a) \includegraphics[width=0.9\hsize]{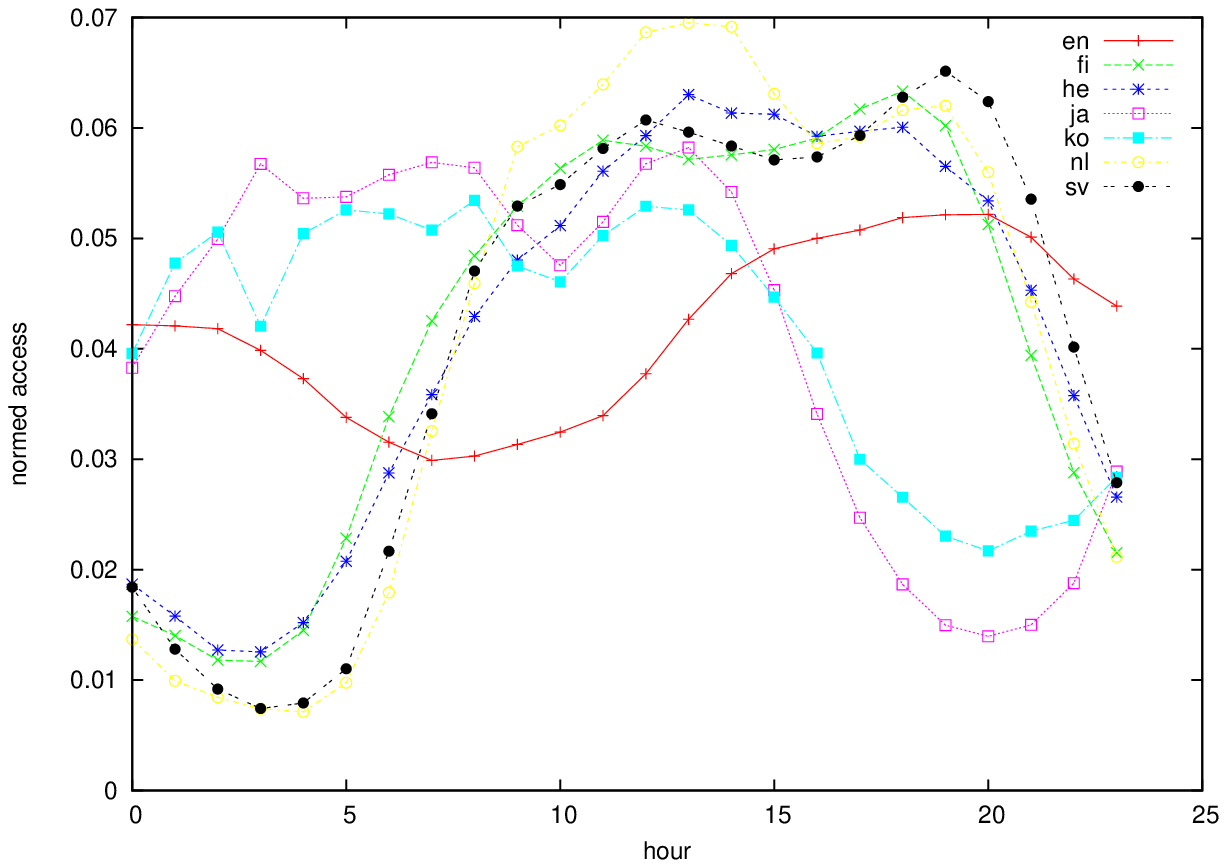}
(b) \includegraphics[width=0.9\hsize]{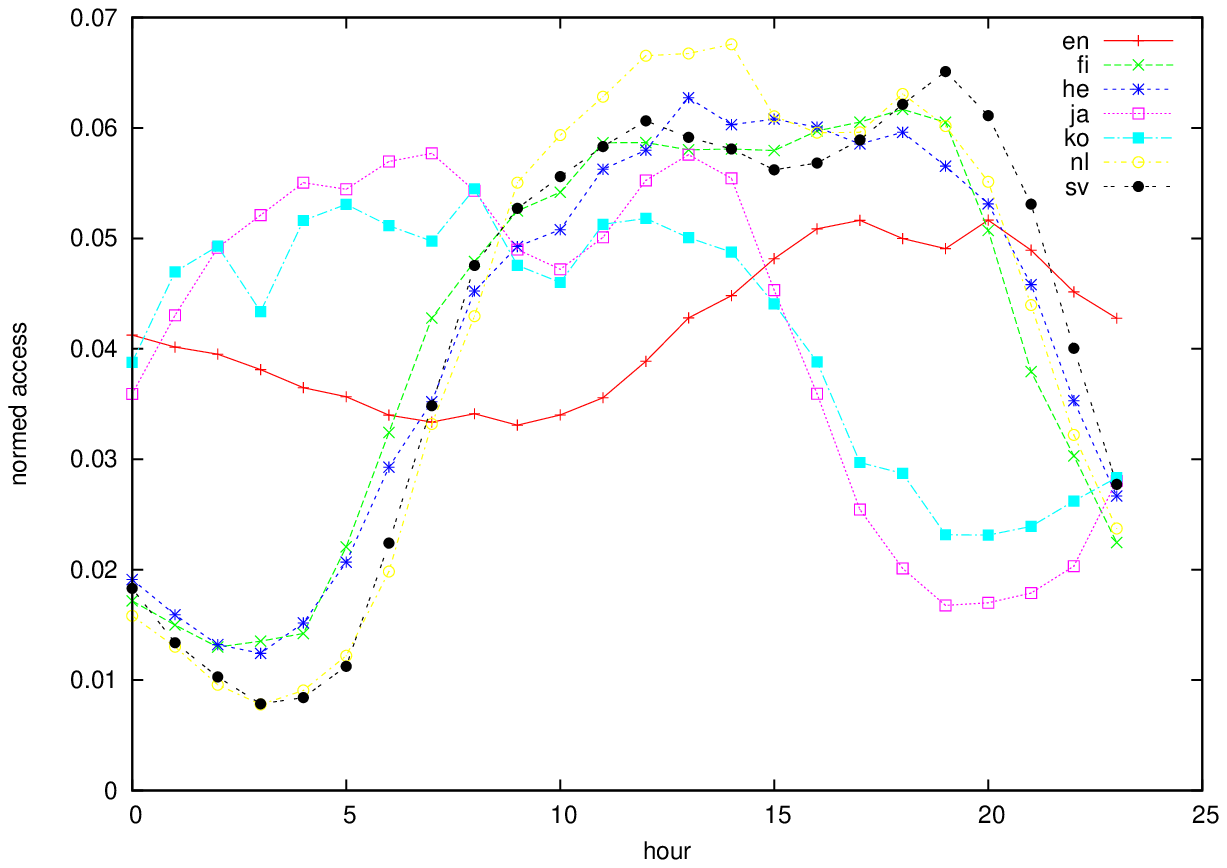}
(c) \includegraphics[width=0.9\hsize]{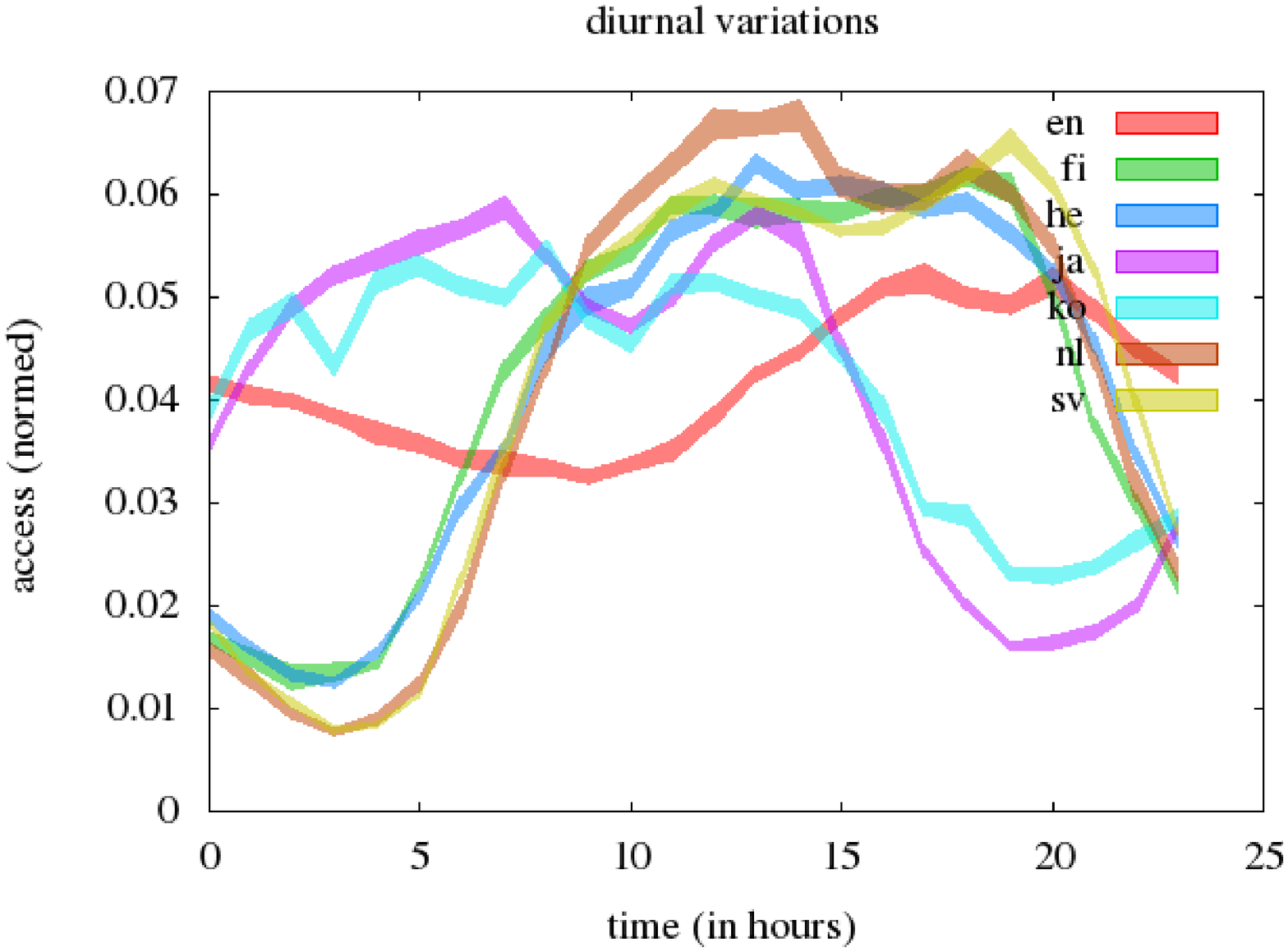}
\caption{Diurnal variations for the different languages with automatically created lists
including (a) only normal articles and (b,c) all pages.  The widths of the bars in (c) 
indicate the single standard deviations of the arithmetic averages over all considered 
pages and confirm that most differences are significant.  Time is reported for UTC time zone.}
\end{figure}

Figure 7 shows that the Wikipedias we chose because they are similar to Swedish (Finish, Dutch, 
and Hebrew) also show similar results for the diurnal variations.  The day-night-cycle is very 
distinctive, and the maxima at lunch time and in the early evenings are similar as well. 
Korean and Japanese day-night-cycle are less distinctive, possibly because of emigrants in the 
US and elsewhere.  Diurnal variations for English are 
very different because of the global context.  Nearly the same results are found for normal 
articles and all pages (namespaces) in group 3.  Figure 7(c) shows that the standard deviation 
for all languages are low, so that the differences between languages are significant in most 
cases.

\subsection{Summertime and Wintertime}

\begin{figure}[t]
(a) \includegraphics[width=0.4\hsize]{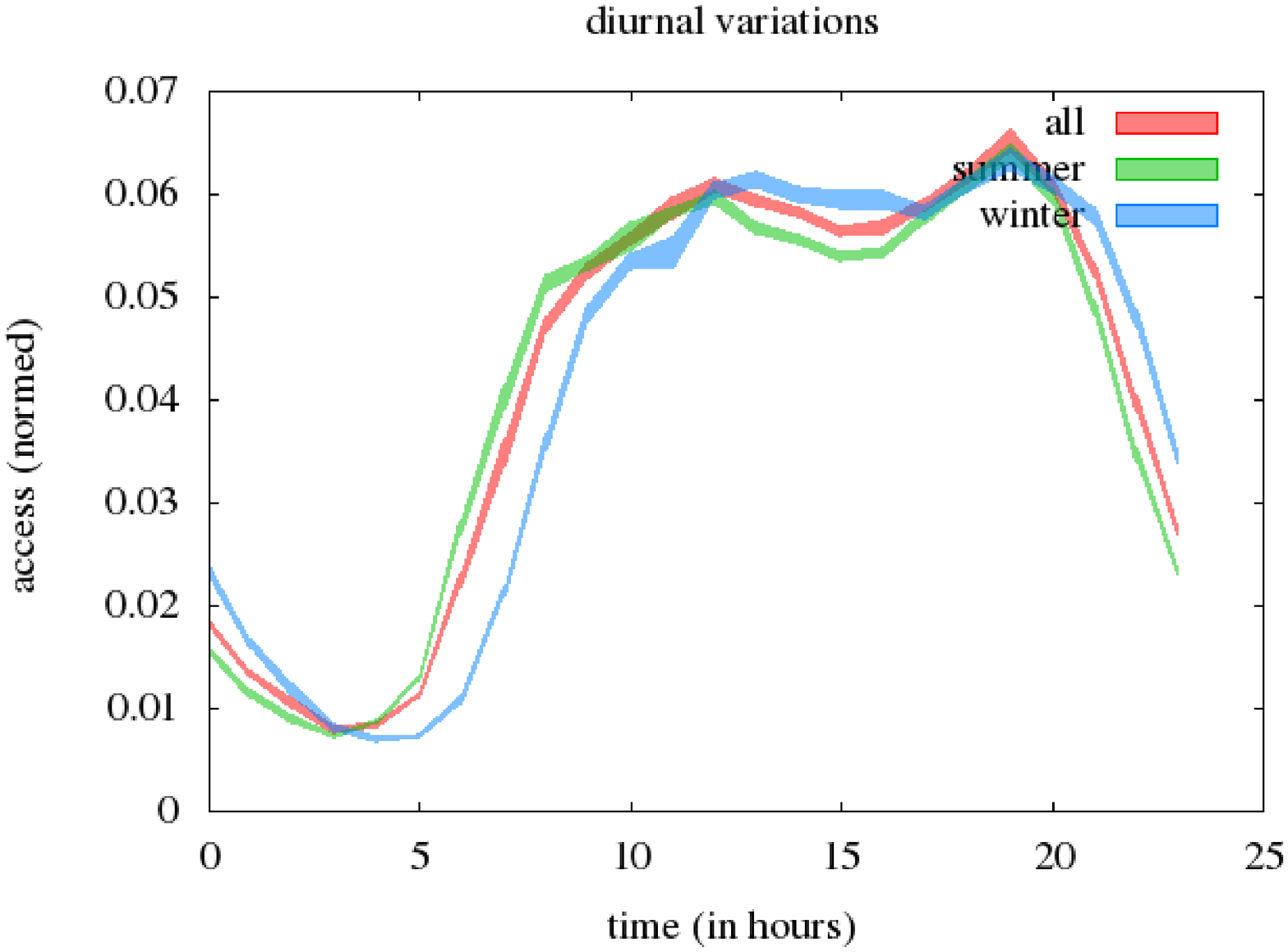}
(b) \includegraphics[width=0.4\hsize]{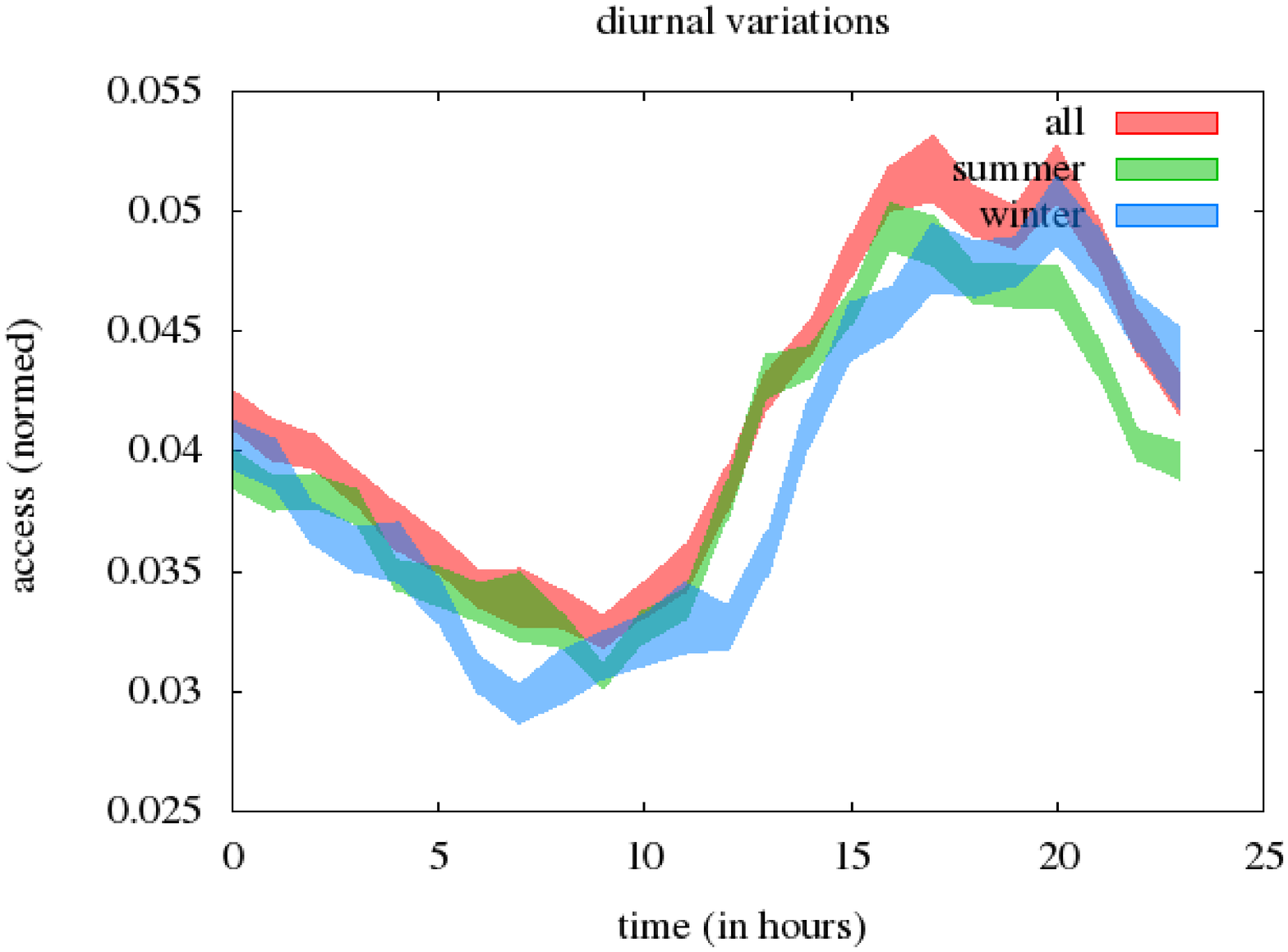} \\ 
(c) \includegraphics[width=0.4\hsize]{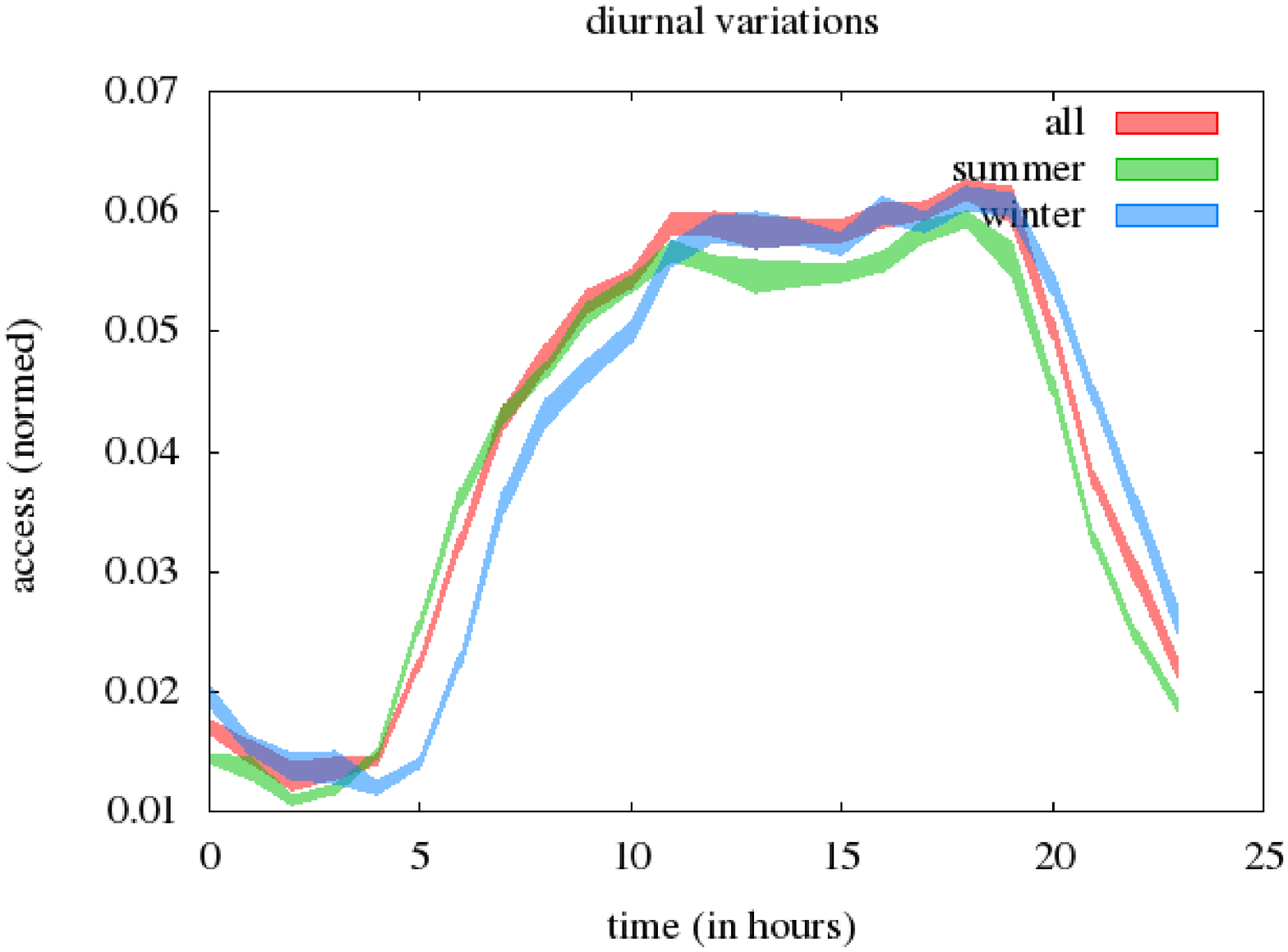}
(d) \includegraphics[width=0.4\hsize]{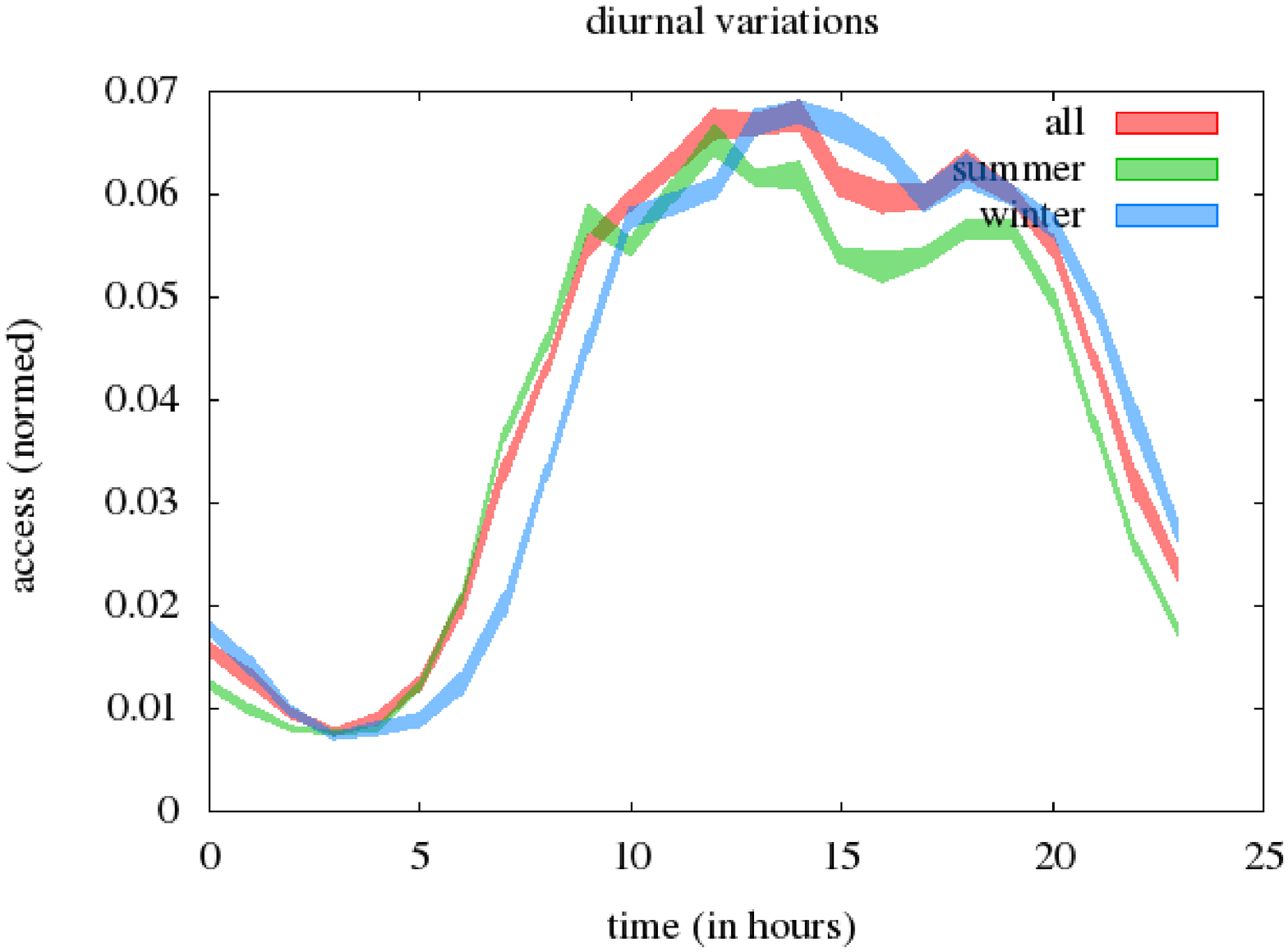} \\
(e) \includegraphics[width=0.4\hsize]{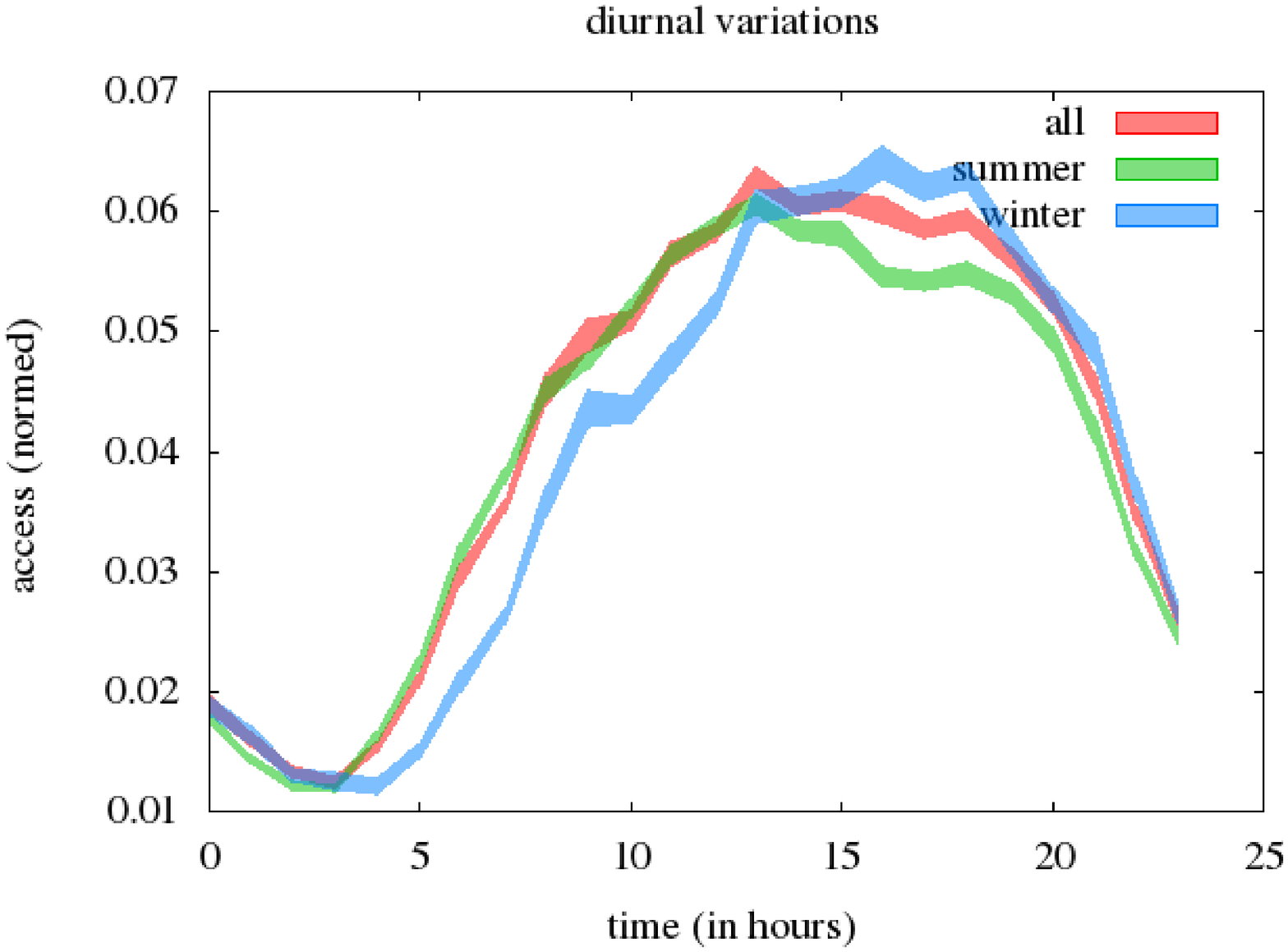}
(f) \includegraphics[width=0.4\hsize]{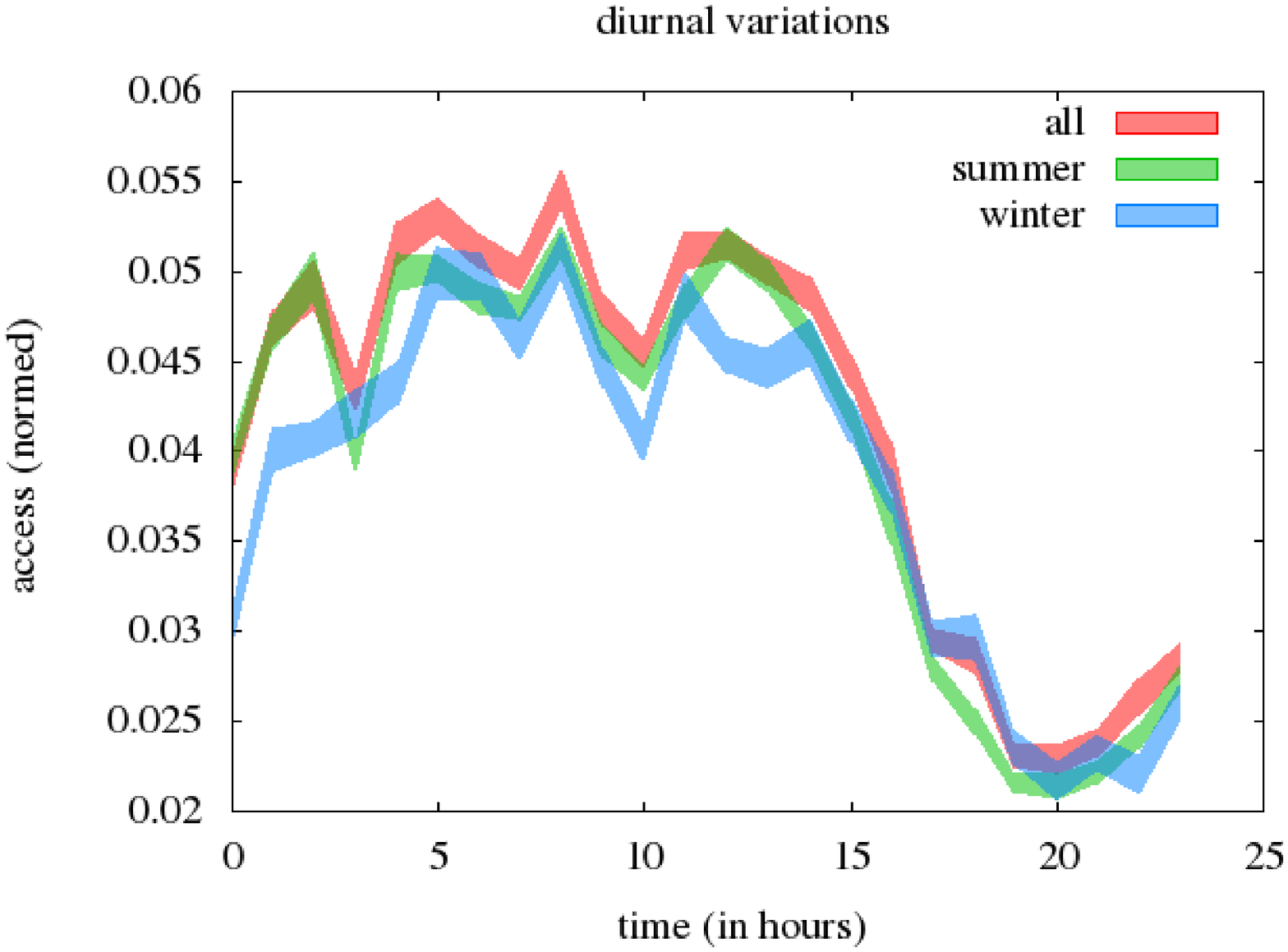} \\
(g) \includegraphics[width=0.4\hsize]{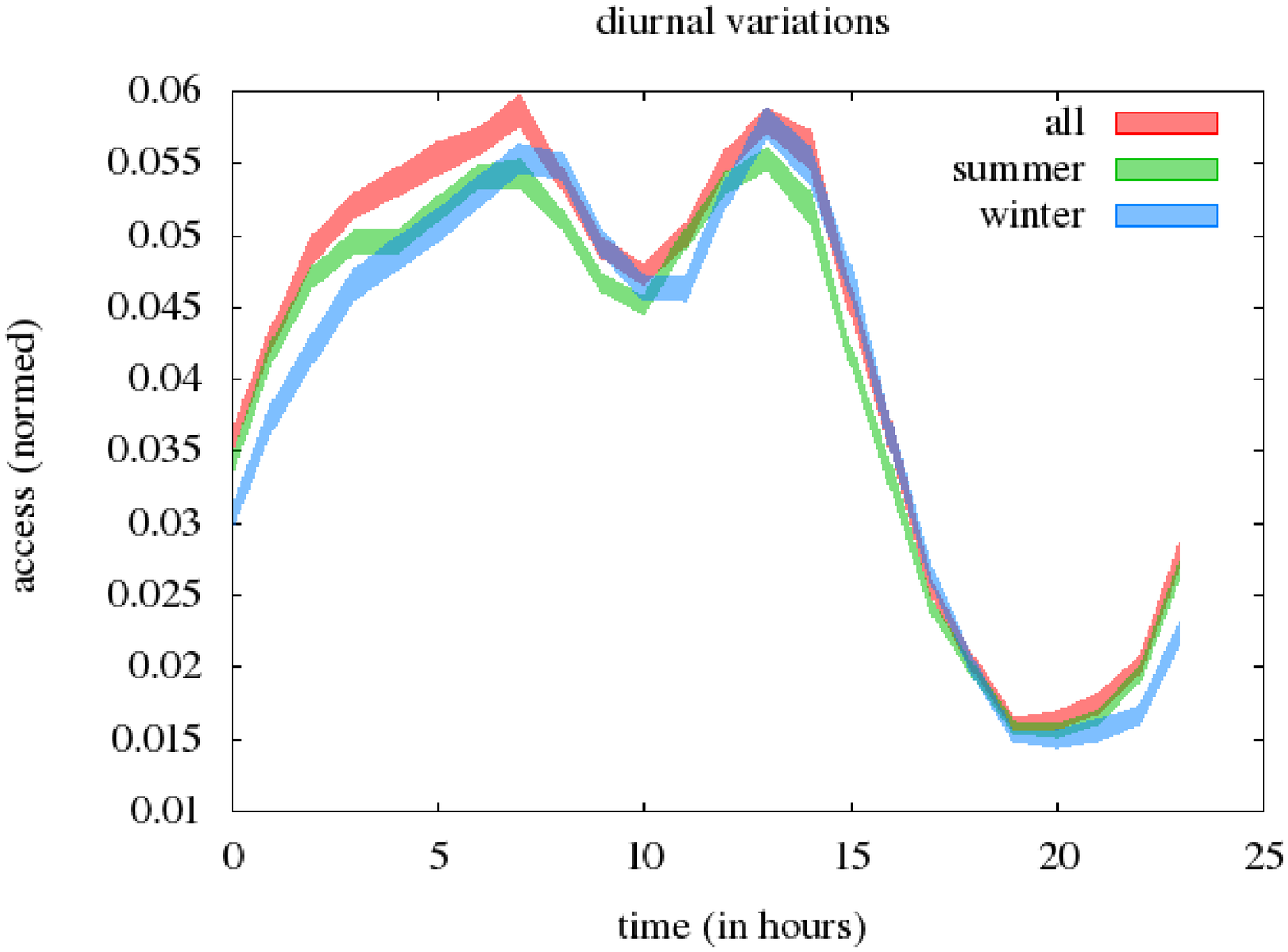}
\caption{Diurnal variations: whole year, summertime and wintertime for (a) Swedish, (b) 
English, (c) Finnish, (d) Dutch, (e) Hebrew, (f) Korean, and (g) Japanese Wikipedia.}
\end{figure}

For Fig.~8 the hours of access where split into two groups: hours of summertime and of 
wintertime.  The different days of change and presence or absence of daylight-saving time 
in each country were taken into account.  The effects of different time zones in Japan and
Korea compared to the European countries are clearly visible.  Interestingly, the two peaks
during daytime are much more pronounced in Japan and practically absent in Israel during 
wintertime.  

\subsection{Intermediate Month in Israel}

\begin{figure}[t]
(a) \includegraphics[width=0.9\hsize]{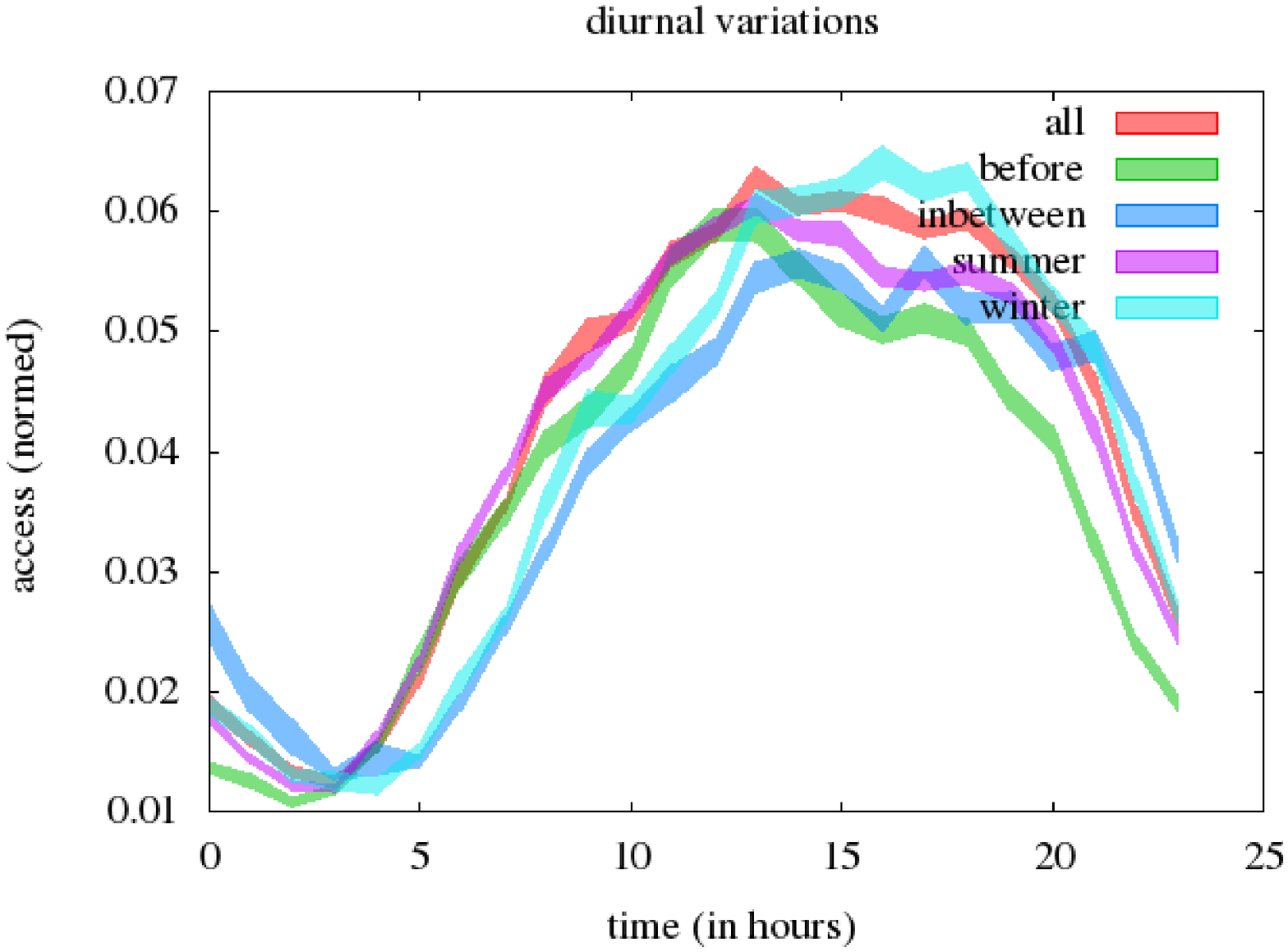}
(b) \includegraphics[width=0.9\hsize]{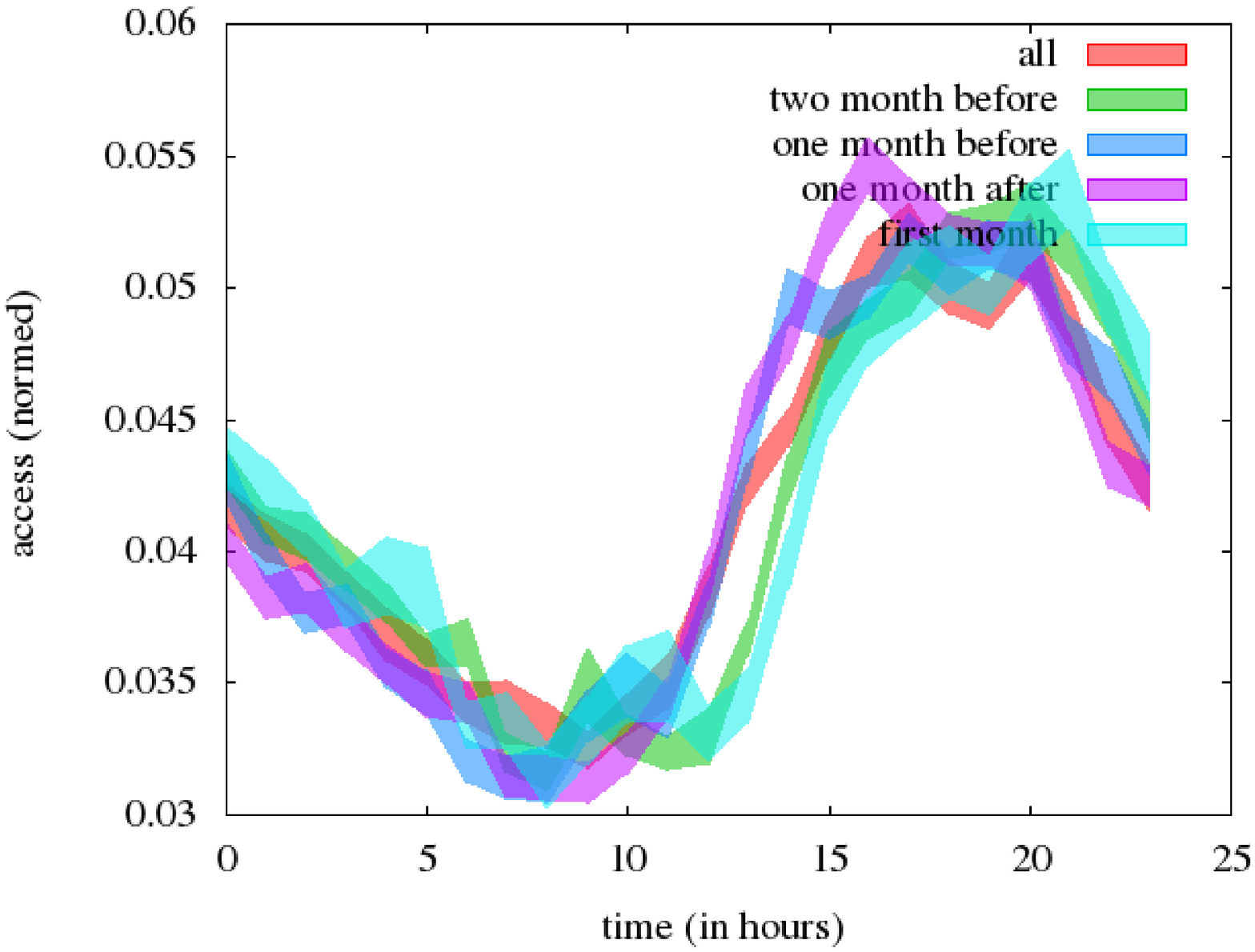}
\caption{Diurnal variations for (a) Hebrew and (b) English Wikipedia in whole year 
(red), last month of daylight-saving time in Israel (green), first month of wintertime 
in Israel, which is last month of daylight-saving time in other countries (blue), 
whole summertime (violet) and wintertime (light blue).}
\end{figure}

Additional information can be obtained from the data of the Hebrew Wikipedia, since the 
change from summertime (daylight-saving time) to wintertime occurred approximately one month 
earlier than in Europe and in most other countries.  This effect can be used to distinguish 
users of Hebrew Wikipedia located in Israel in other parts of the world.  Although the changes
seen in Fig. 9(a) do not appear to be fully systematic, one can see that the behavior during 
the first month of wintertime (blue) is much more similar to the other months of wintertime
(light blue) than to the last month of summertime (green).  This indicates that most users 
of the Hebrew Wikipedia are actually located in the land of Israel and not in Europe or the 
US.  We have also tried to identify changes in access volume for the English Wikipedia this
way to see if people in Israel are using it, see Fig. 9(b).  However, the fraction of Israeli 
users of the English Wikipedia is too small compared with other users, so that changes due to 
Israeli daylight-saving time are not significant in English Wikipedia access volumes. 
Random variations from month to month turned out to be larger than these effects. 

\subsection{Linear Combination Analysis}

In another analysis step we have used two functions of diurnal variations as templates and 
linearly combined them to express a third one.  This procedure corresponds to the assumption 
that, e.g., the diurnal variation of Hebrew Wikipedia usage during the first month of wintertime
in Israel is composed of users from Israel (who approximately follow the typical Hebrew
wintertime pattern) and users from other countries (who approximately follow the typical Hebrew
pattern for the last month of summertime).  The result is that a coefficient of only 0.011
appears in front of the last-month-of-summertime pattern (foreign users), while 0.928 appear
in front of the wintertime pattern (users in Israel).  This results confirms that most users
of the Hebrew Wikipedia are located in Israel and shows that meaningful results can be obtained 
from such approximation tests.

\begin{table}[b]
\caption{Fits of Swedish Wikipedia access rate diurnal variations by two templates:  first 
template as given in column 1 with time delay according to column 2 and second template 
English Wikipedia access rate diurnal variations.  The resulting coefficients of the linear
combination are given in columns 3 and 4.}
\begin{tabular}{c|c|c|c}
first template & delay & first coeff. & English coeff. \\ \hline
Finish & 1h & 1.15 & $-0.16$ \\
Dutch & 0h & 0.89 & 0.12 \\
Japanese & 8h & 1.31 & $-0.30$ \\
Hebrew before & 1h & 1.21 & $-0.06$ \\
Hebrew summer & 1h & 1.23 & $-0.19$ \\
Hebrew winter & 1h & 1.10 & $-0.07$ \\
Hebrew between & 1h & 1.34 & $-0.22$
\end{tabular}
\label{Tab:1}
\end{table}

Therefore we tried to express the diurnal variations of Swedish Wikipedia access rates by 
various templates including the English diurnal variations.  The goal was to identify the 
relation between English and Swedish Wikipedia in comparison with other Wikipedias.  Table 1
shows that the corresponding coefficients for the English diurnal variations' template are 
negative in most cases.  This indicates that the approach is not sufficiently reliable.

\section{Diurnal Variations of Edit Rates in Group 3}

In this Section the diurnal variations of edit number and edit rates are studied for each 
of the seven languages in group 3 and compared with diurnal variations of access rate. 

\begin{figure}[t]
\centering \includegraphics[width=0.9\hsize]{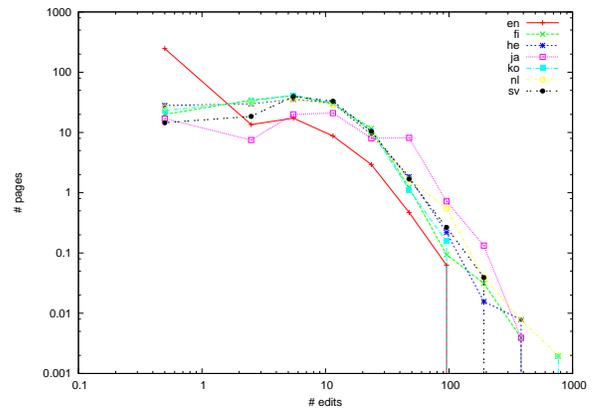}
\caption{Histogram of the numbers of pages in group 3 versus the specific number of edits
for all seven considered languages (see legend). }
\end{figure}

\begin{figure}[t]
(a) \includegraphics[width=0.9\hsize]{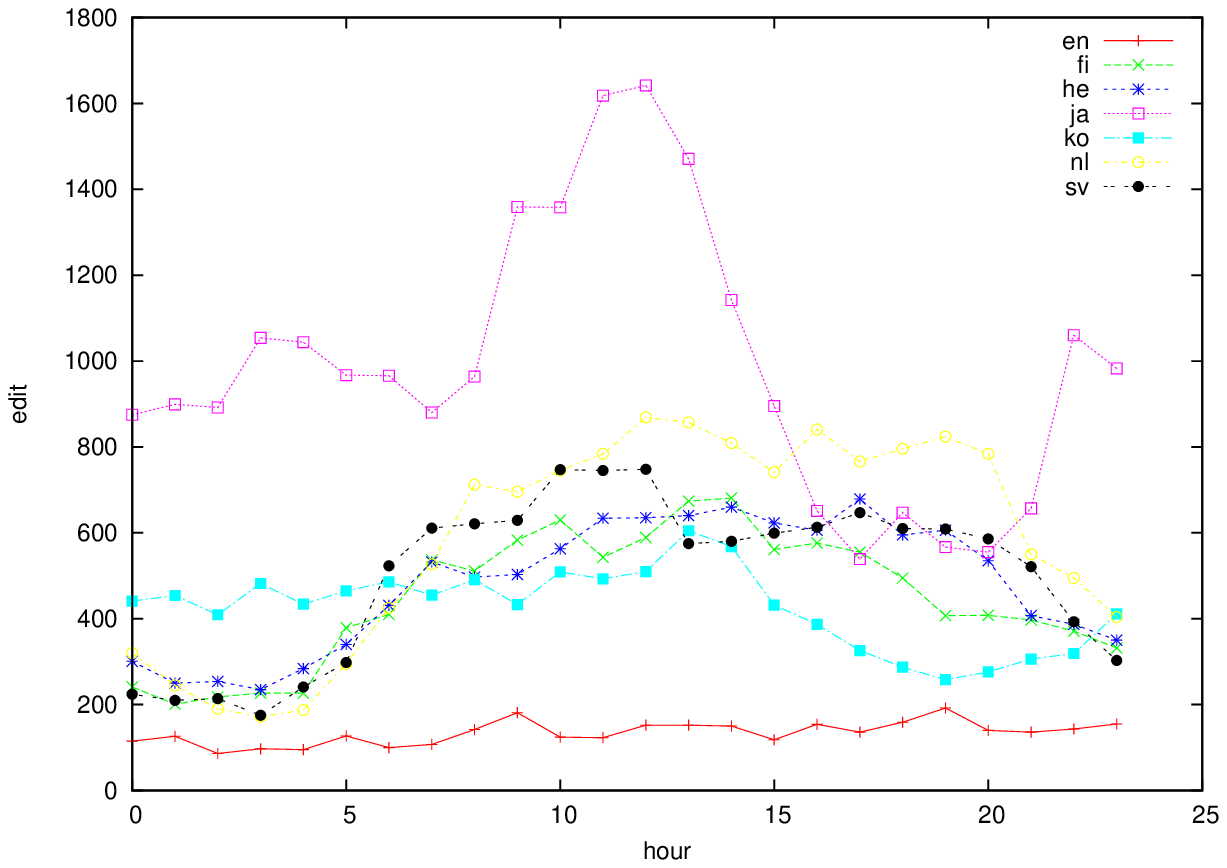}
(b) \includegraphics[width=0.9\hsize]{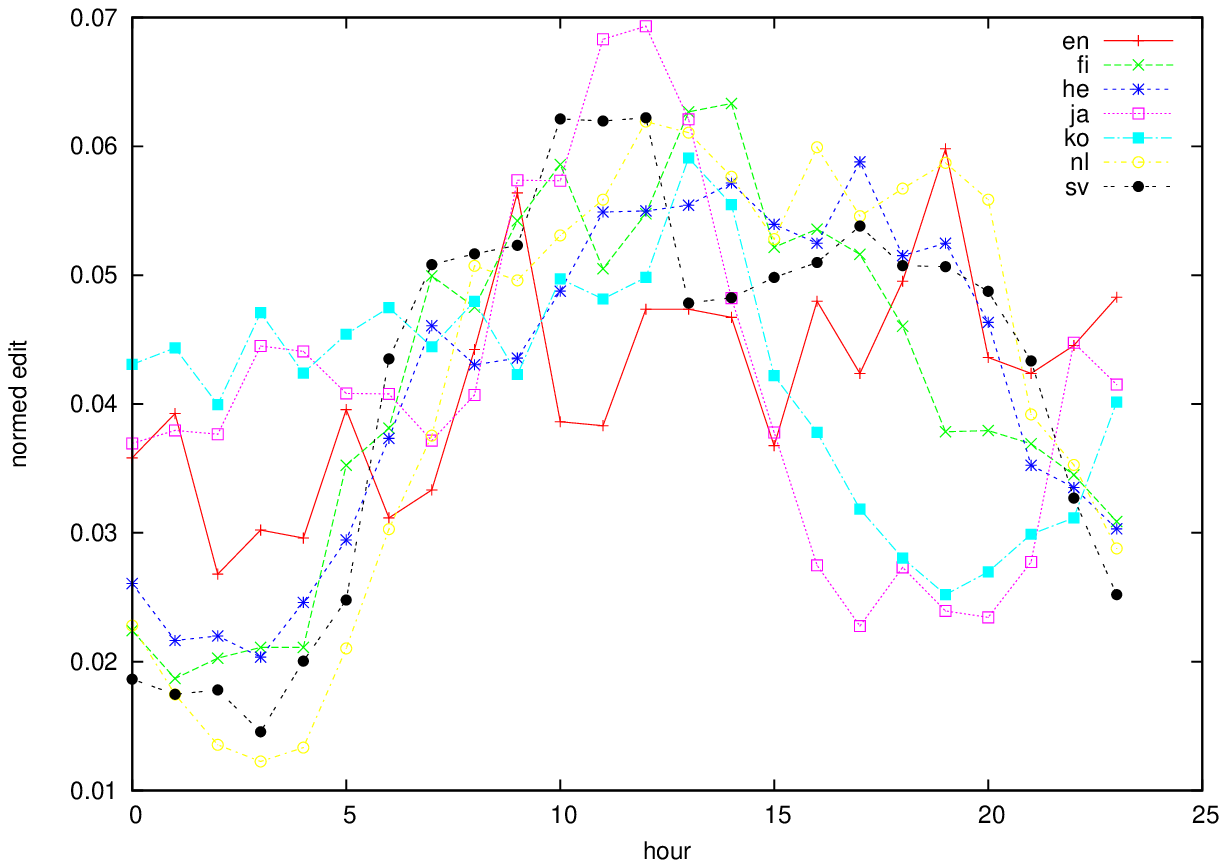}
\caption{Diurnal variations of edits activity for the seven different languages. 
(a) Total and (b) relative number of edits per hour for all pages in group 3.  See 
legends for country codes.}
\end{figure}

Figure 10 shows that the editorial activity is quite similar for the considered group of
Wikipedia articles.  For large numbers of edits one can see a power-law decrease of the 
number of articles according to $N(\epsilon) \sim \epsilon^{-\gamma}$ with $\gamma \approx 
2.7$ with $\epsilon$ the total number of edits in our observation.  English articles are 
edited somewhat less frequently in our observation period, while Japanese articles are 
edited somewhat more frequently. Swedish articles are not deviating from most other languages
regarding the edit frequency distribution.  This results is somewhat surprising. 
It may be that many of the English articles were already in good shape before 2009.  The 
English Wikipedia might also have more short articles compared to the number of editors or 
average article length.  On the other hand, Swedish editors are usually much more active 
than, e.~g. Koreans or Germans, and Swedish articles tend to be usually a bit shorter. 

\begin{figure}[t]
(a) \includegraphics[width=0.4\hsize]{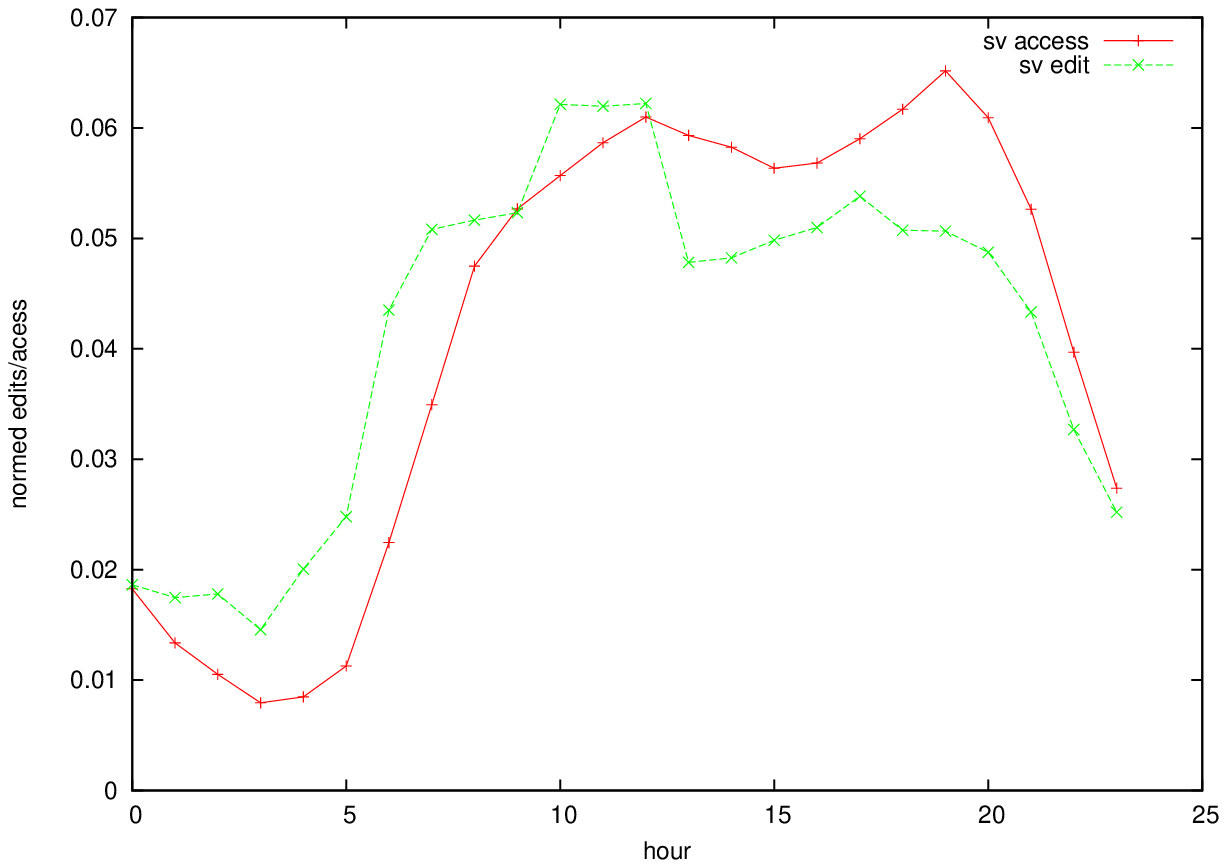}
(b) \includegraphics[width=0.4\hsize]{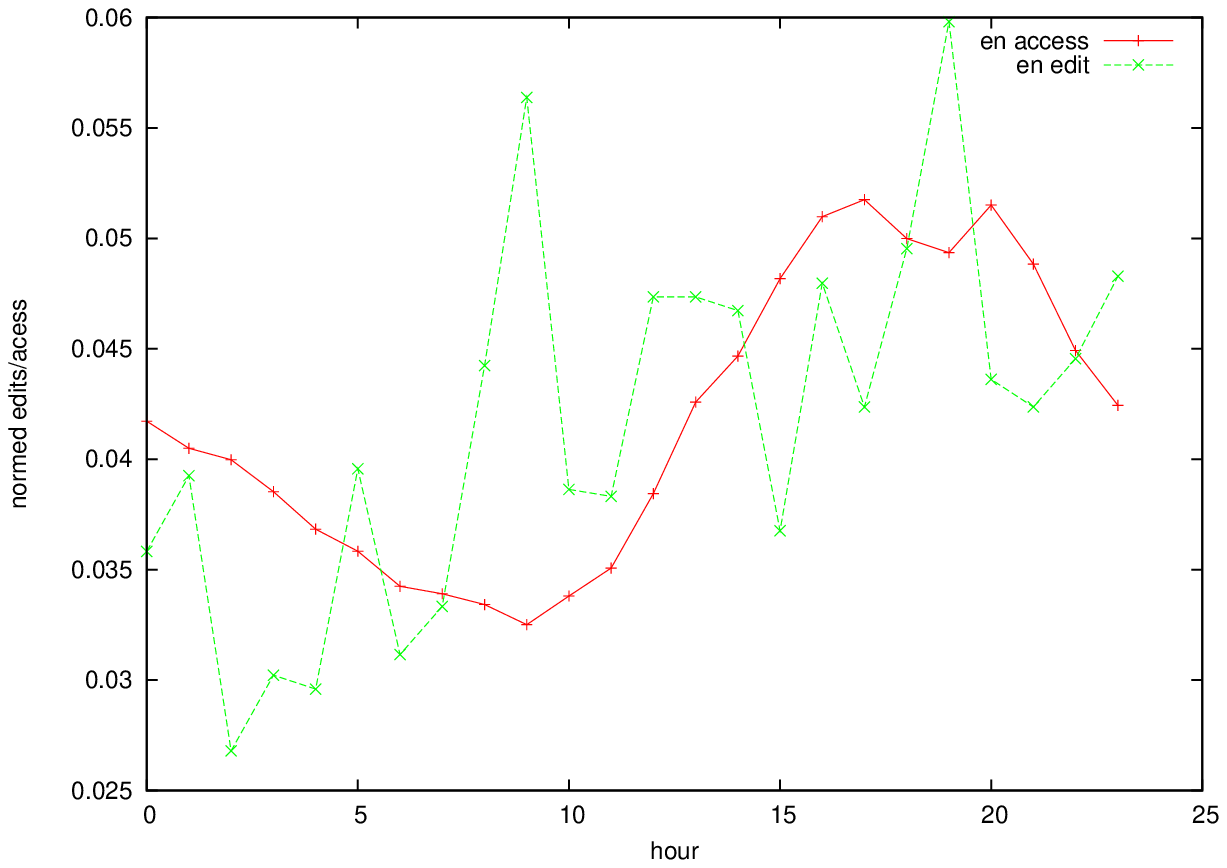} \\ 
(c) \includegraphics[width=0.4\hsize]{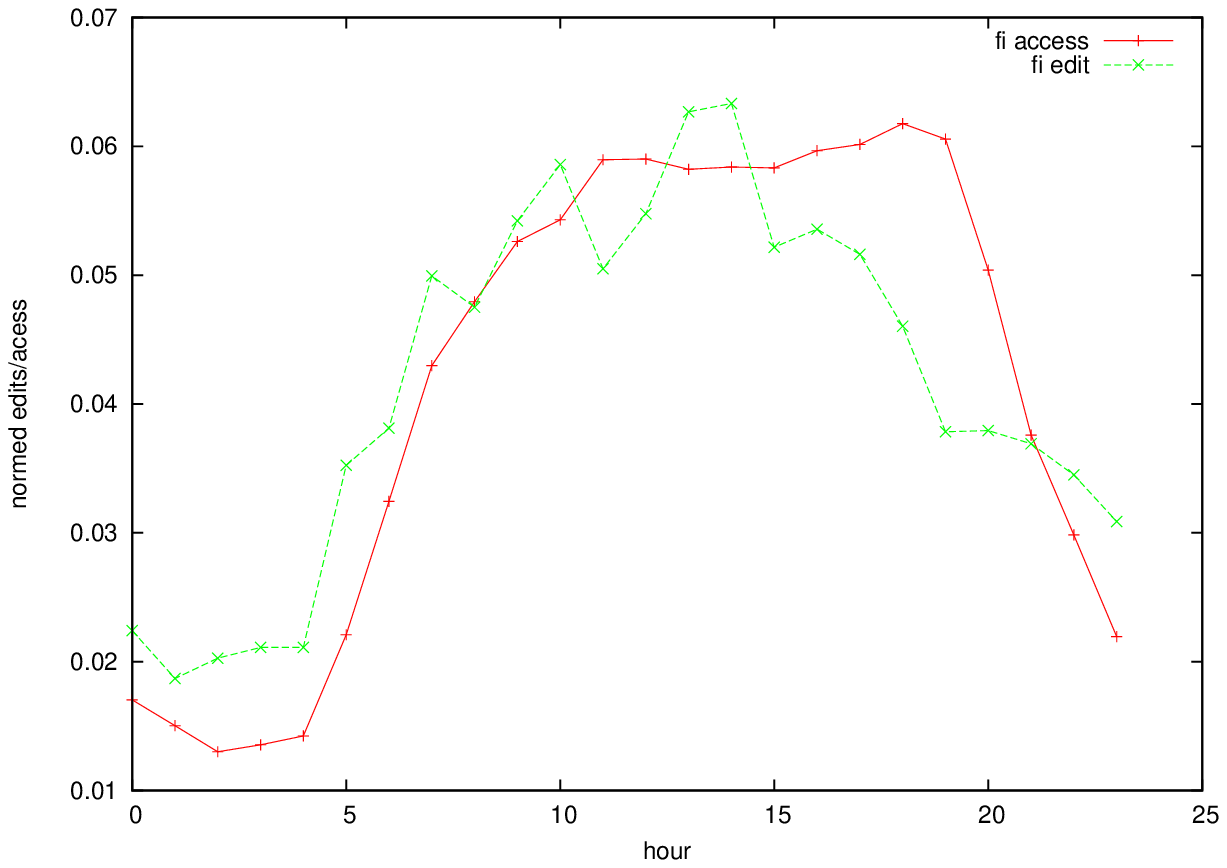}
(d) \includegraphics[width=0.4\hsize]{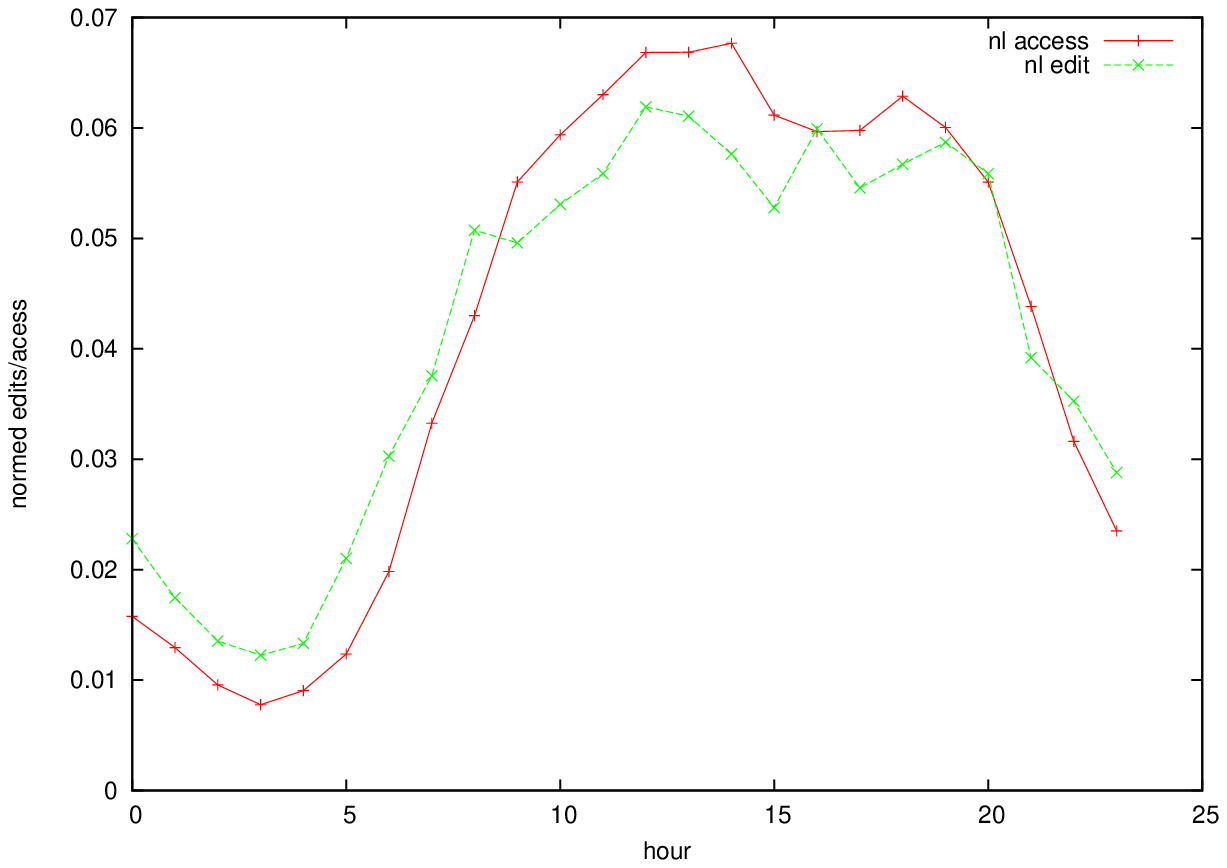} \\
(e) \includegraphics[width=0.4\hsize]{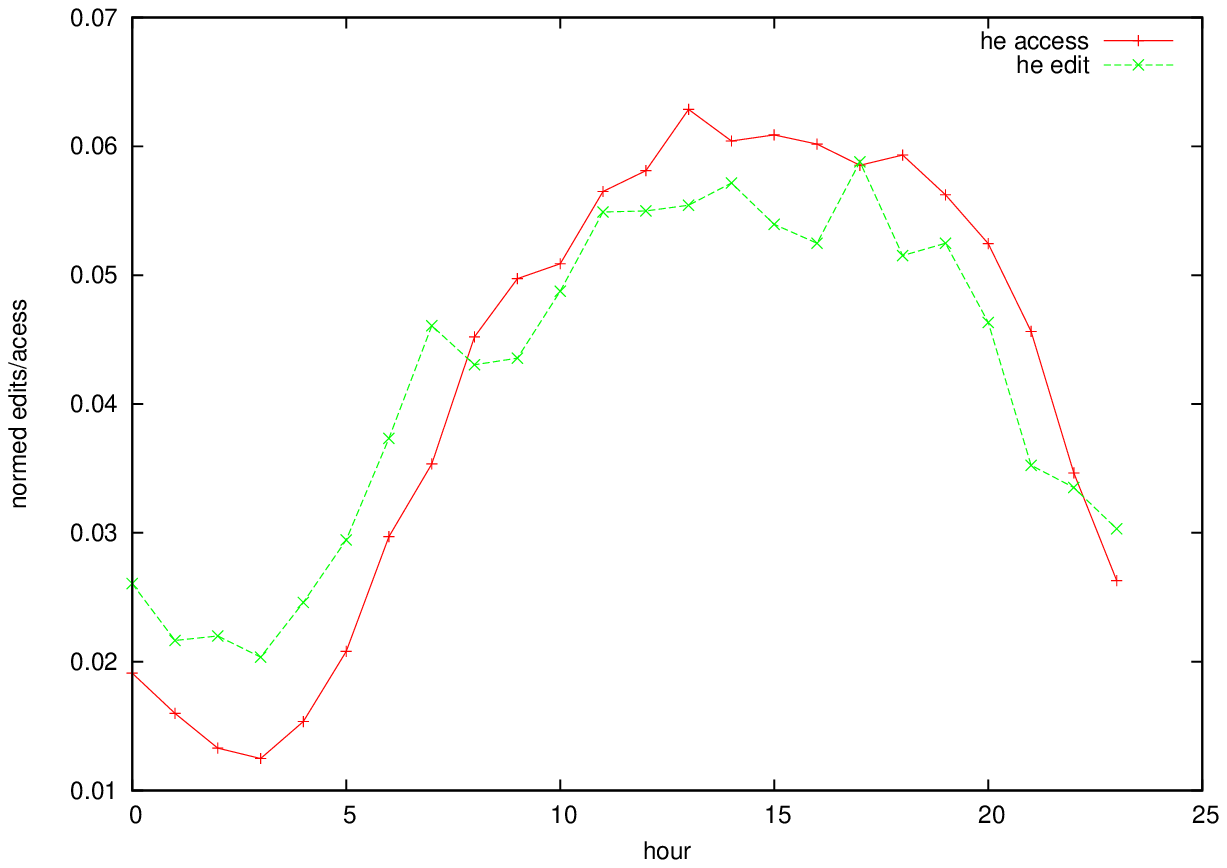}
(f) \includegraphics[width=0.4\hsize]{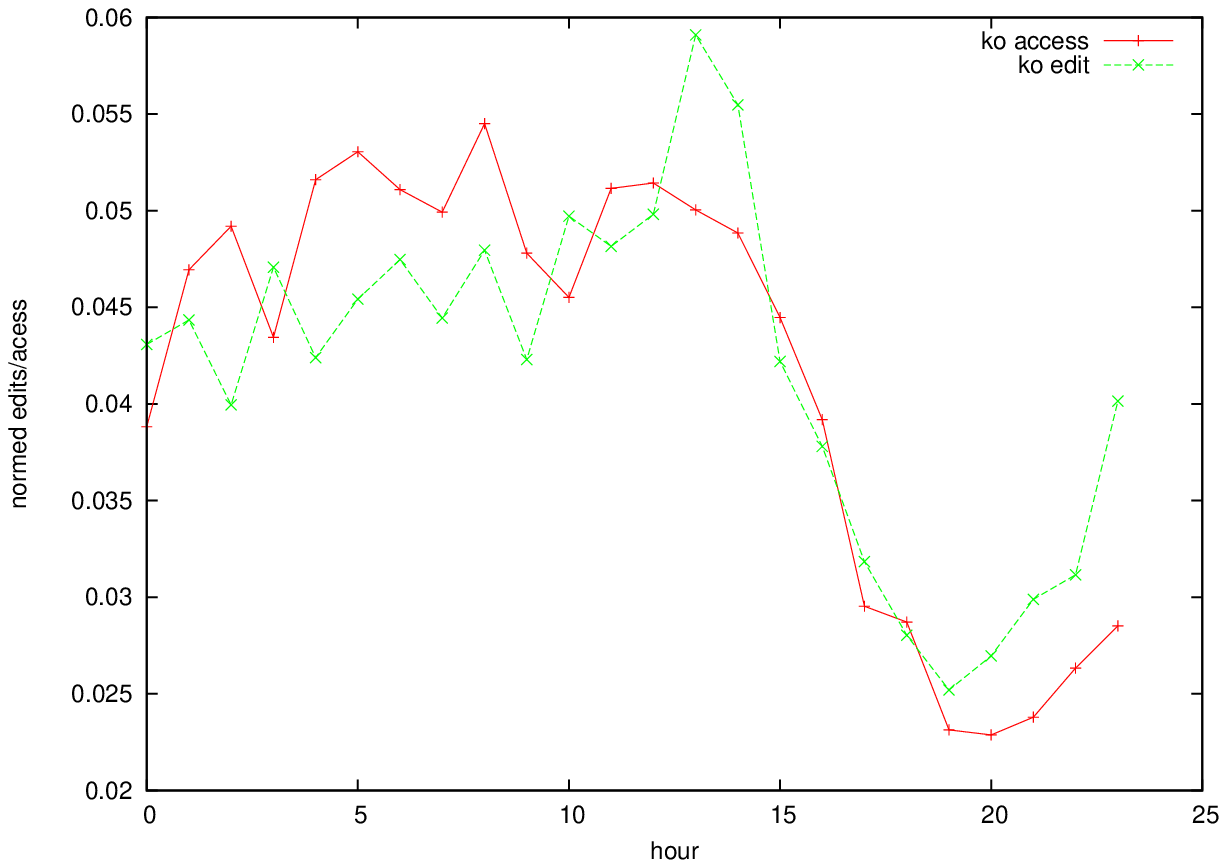} \\
(g) \includegraphics[width=0.4\hsize]{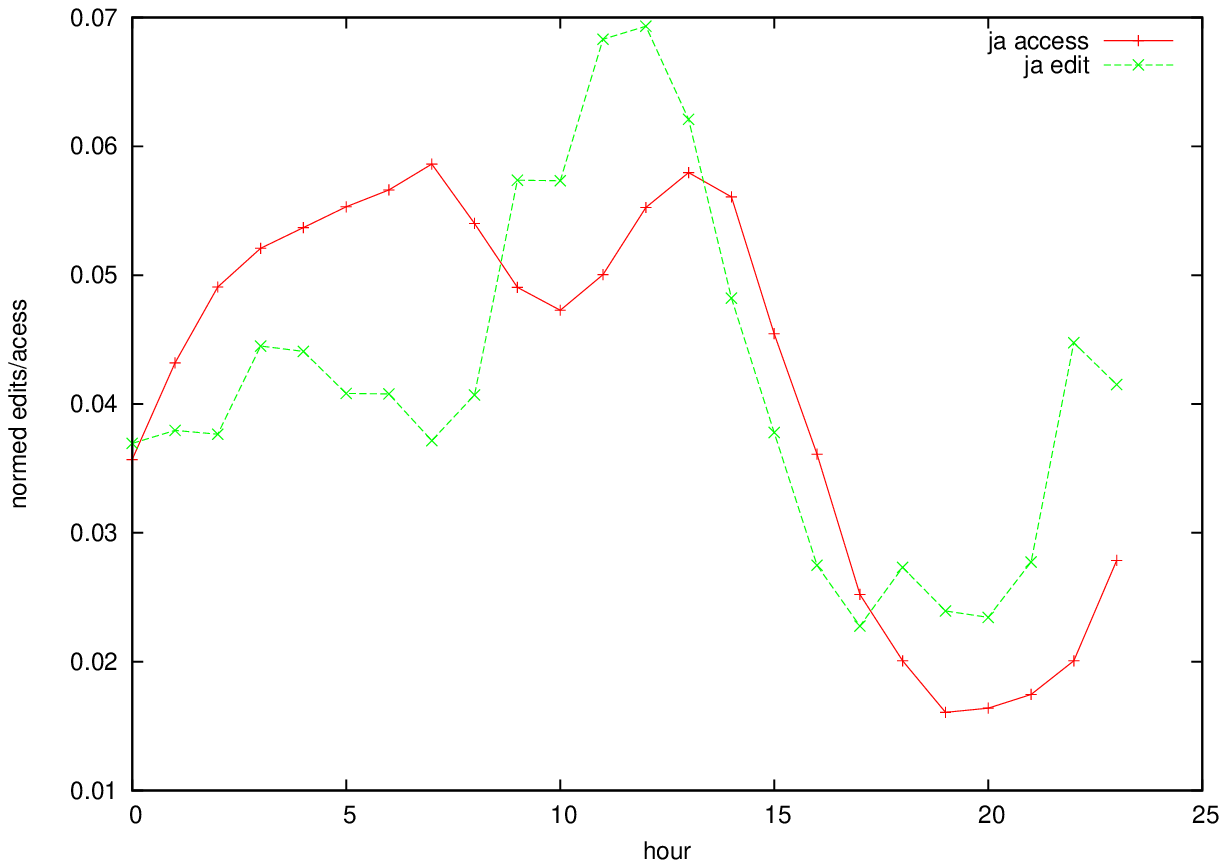}
\caption{Normalized diurnal variations of access rate (red) and edit rate (green) 
for (a) Swedish, (b) English, (c) Finnish, (d) Dutch, (e) Hebrew, (f) Korean, and 
(g) Japanese Wikipedia. The data for English Wikipedia (part (b)) is a bit more noisy, 
since edits are less frequent in general (see Fig. 11(a)); furthermore, the diurnal 
variation is less, so that the vertical axis is stretched here.}
\end{figure}

Figure 11 compares the total and the normalized diurnal variations of edit activity for 
all seven languages.  Again, one can see that total edit activity is less in the English
articles and more in the Japanese articles.  Normalized edit activity variations are much 
more noisy than access volume variations (see also Fig. 7), since statistics are much 
worse for edit activity.  Nevertheless, Fig. 12 shows significant similarities in the
diurnal variation patterns.  In all cases, the diurnal curve of edit activity runs ahead
of the curve for access volume.  This indicates that editing Wikipedia pages is usually
done earlier than viewing pages.  For Swedish Wikipedia (Fig. 12(a)), the morning peak of
editorial activity is much more pronounced than the afternoon peak.  In addition, the 
minimum during the night is twice as large for editorial activity than for access activity
(see also Fig. 12(a)).  Since typical Wikipedia editors are students with no real regulated 
work time, editors are probably much more night active than average readers.  However, we 
cannot exclude that the editing of Swedish Wikipedia pages is more often done from abroad 
than the viewing of Swedish Wikipedia pages.  In general, the diurnal
variation pattern of editorial activity are too noisy to allow attributing editors to 
different time zones just based on these patterns in a quantitative way.

\begin{figure}[b]
\centering \includegraphics[width=0.8\hsize]{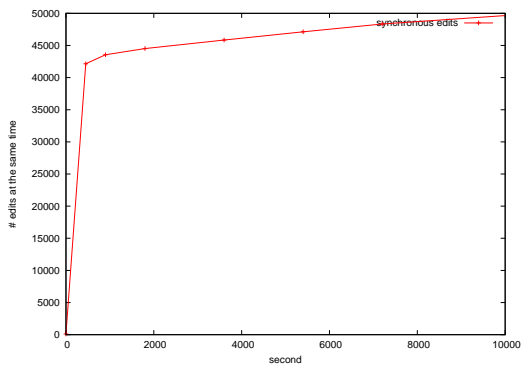}
\caption{Number of synchronous edits in corresponding Swedish and English Wikipedia articles
in group 2 versus allowed time difference between the two events in seconds.  At 900 seconds 
there is a saturation.  The linear increase for larger time differences comes from randomly 
synchronous edits.}
\end{figure}

\begin{figure}[t]
(a) \includegraphics[width=0.75\hsize]{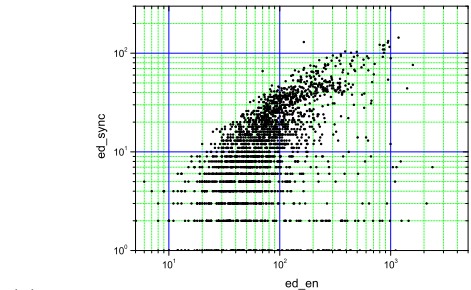} \\
(b) \includegraphics[width=0.75\hsize]{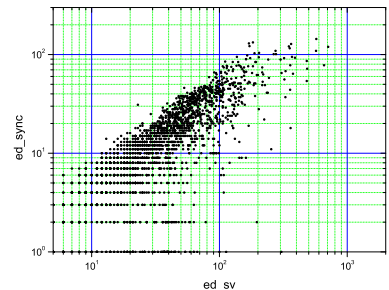} \\
(c) \includegraphics[width=0.75\hsize]{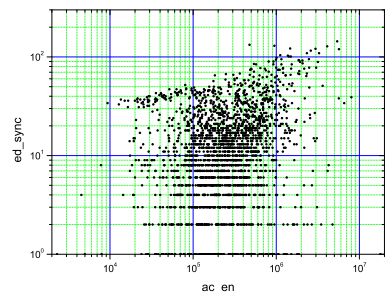} \\
(d) \includegraphics[width=0.75\hsize]{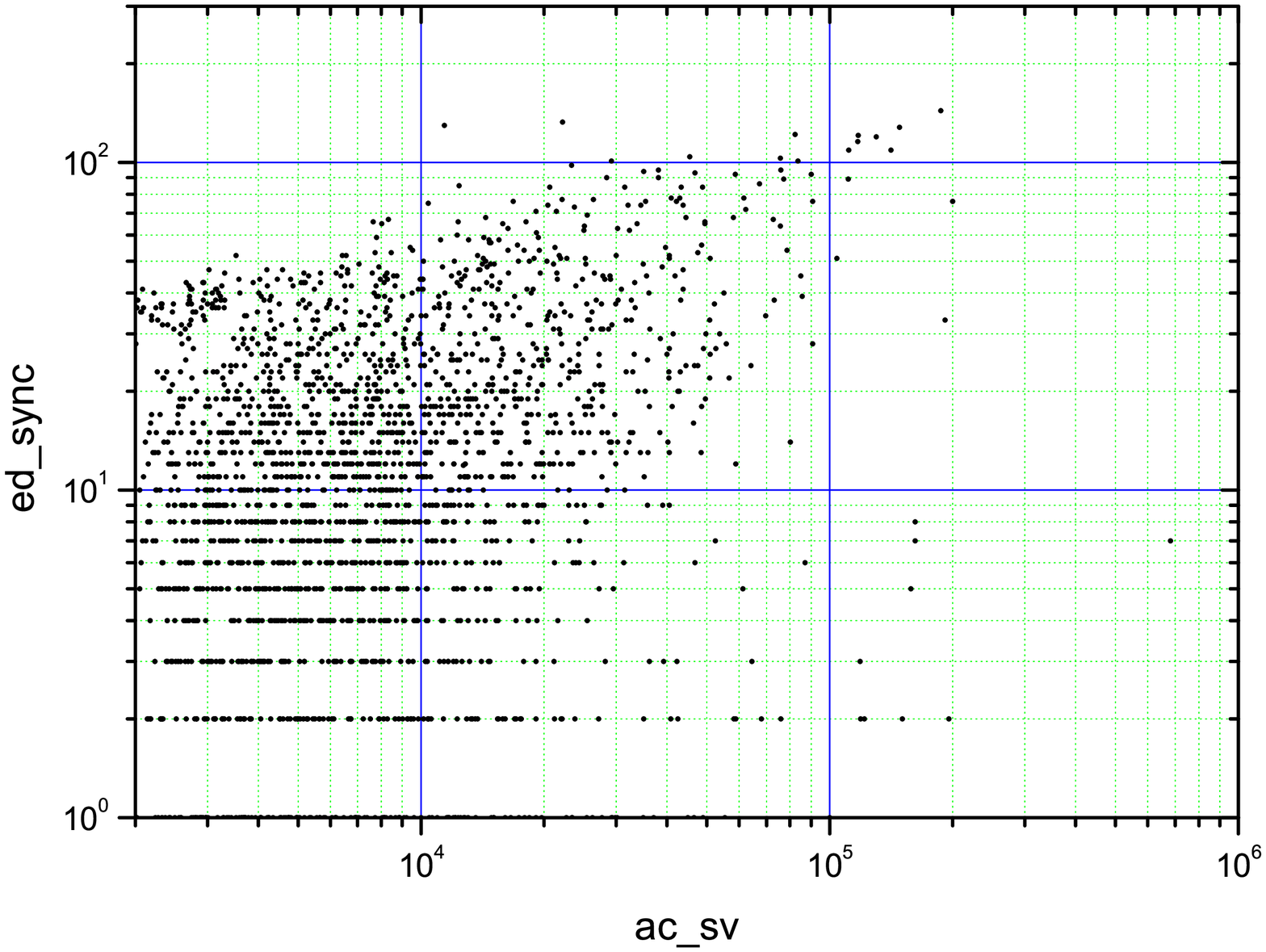}
\caption{Number of synchronous edits of Swedish and English articles in group 2 versus 
(a,b) total number of edit events and versus (c,d) total number of access events for 
the (a,c) English and (b,d) Swedish article.}
\end{figure}

\section{Synchronous Edits in Different Language Versions}

In this Section we study the occurrence of synchronous edit events in Swedish and English
article pairs from group 2.  A synchronous edit event means that both language versions 
of an article (with corresponding topic) were modified simultaneously.

Figure 13 shows the total number of synchronous edits versus the permitted time difference.  
Clearly, no edits occur exactly at the same time, and if larger time differences are allowed, 
more edits are considered synchronous.  Randomly synchronous edit events cause a linear 
increase for large time differences.  To include as few random events as possible but still 
identify most of the practically synchronous edit event, we have chosen a threshold of 15 
minutes for the definition of synchronous events based on Fig. 13.  

Figure 14 shows the resulting number of synchronous edit events versus the total number of
edits and the total number of accesses of the Swedish and English version of the article.
In Figs. 14(a,b) one can see that there are article pairs where nearly all edits are 
synchronous (close to the diagonal), while for others there are only very few (or even
none, not shown) synchronous edits.  However, no clear clustering of different article 
subgroups is observed.  However, Figs. 14(c,d) show that there are clusters of articles with
particular frequent synchronous edits (30-50 times in the considered time period), but 
fairly low total access volumes.  These article clusters might be related with edit wars:
there are quite active editors with a special focus.

\section{Comparison of Growth of Article Number and Article Quality}

Several models are used to describe the growth process of networks.  Two very popular network 
models are the random graph model and the scale free network.  Both models describe how the 
internal structure evolves in time, based on the degree distribution.  In the first case, 
one assumes that all nodes (pages) already exist, and the growth process consists of adding 
links, one in each time step.  In the second case, one page is added in each time step.  This 
means a new link and a new page are created at the same time.

In a real network, like the Wikipedia content network, both processes of adding new pages and 
adding new links between pages are coupled and cannot be separated from each other.  In order 
to describe the growth of the Swedish Wikipedia project in more detail we analyze the growth 
rates for the number of pages and the number of links.  Because we have several types of 
links we also compare the growth rates of the number of links for those types.  {\it Internal
links} are links within the same Wikipedia (same language) and redirects to another page of
the same Wikipedia.  Internal links represent semantic relations between the terms the pages 
are about or just relations between topics or concepts which are used within a certain page.
If the meaning of a term is ambiguous, special pages help to show users all possible meanings 
(based on other pages).  Such pages do not contribute much text, but this structural information 
is of a high value and increases the usability of Wikipedia.  {\it External links} are links 
to another language (Interwiki links) and links to pages outside the Wikipedia project (e.g., 
references).  The frequency of such links represents an important quality indicator for 
Wikipedia articles.

\subsection{Evolution of the Degree Distribution}

\begin{figure}[t] \label{fig81}
\includegraphics[width=0.9\hsize]{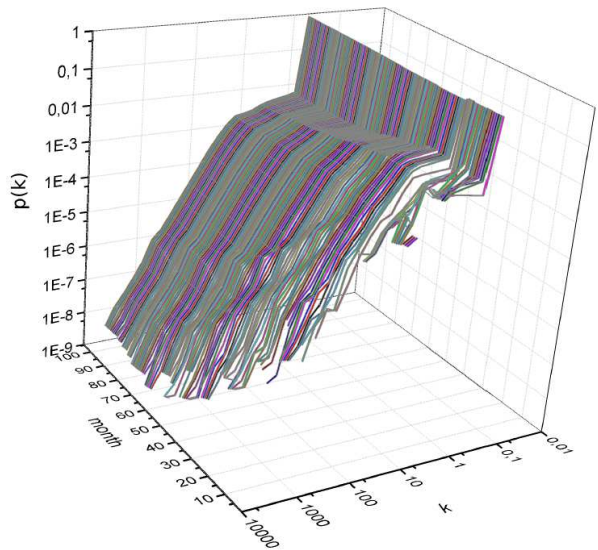}
\caption{Degree distribution $p(k)$ (i.e., distribution of the number $k$ of links per page) 
for internal links in the Swedish Wikipedia project. One curve is shown for each month from 
January 2001 till December 2009.}
\end{figure}

All links contribute to the Wikipedia's structure which evolves over time.  The creation of a 
new link is a result of an edit activity of an user.  Figure 15 shows the temporal evolution of
the internal link degree distribution for all pages of the Swedish Wikipedia.  Redirects and 
external links are disregarded in this plot.  Already since the beginning in 2002 the degree 
distribution can be described by a power law, with the exception of pages with a very low 
degree (low number of links).  While pages are added over time, the distribution changes and its 
power-law shape becomes more obvious, since the range of degrees becomes wider.  Actually, most 
of the pages have much more than ten internal links and are well described by a power-law
degree distribution.  Only the number of pages with less than ten internal links is smaller 
than assumed in the scale-free model that predicts power-law degree distributions.

\subsection{Growth of the Content Network and Structural Changes}

\begin{figure}[t]
(a) \includegraphics[width=0.9\hsize]{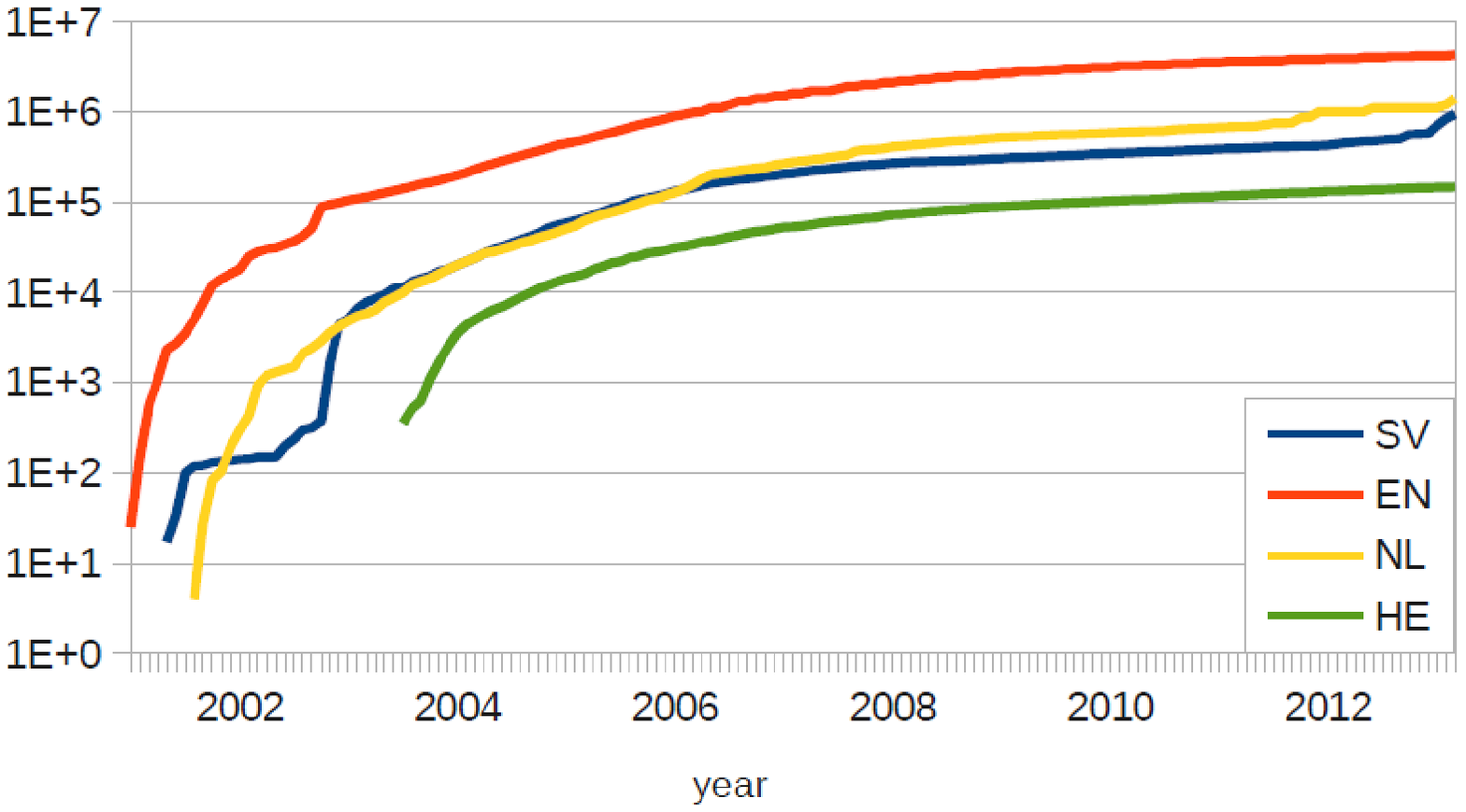}
(b) \includegraphics[width=0.9\hsize]{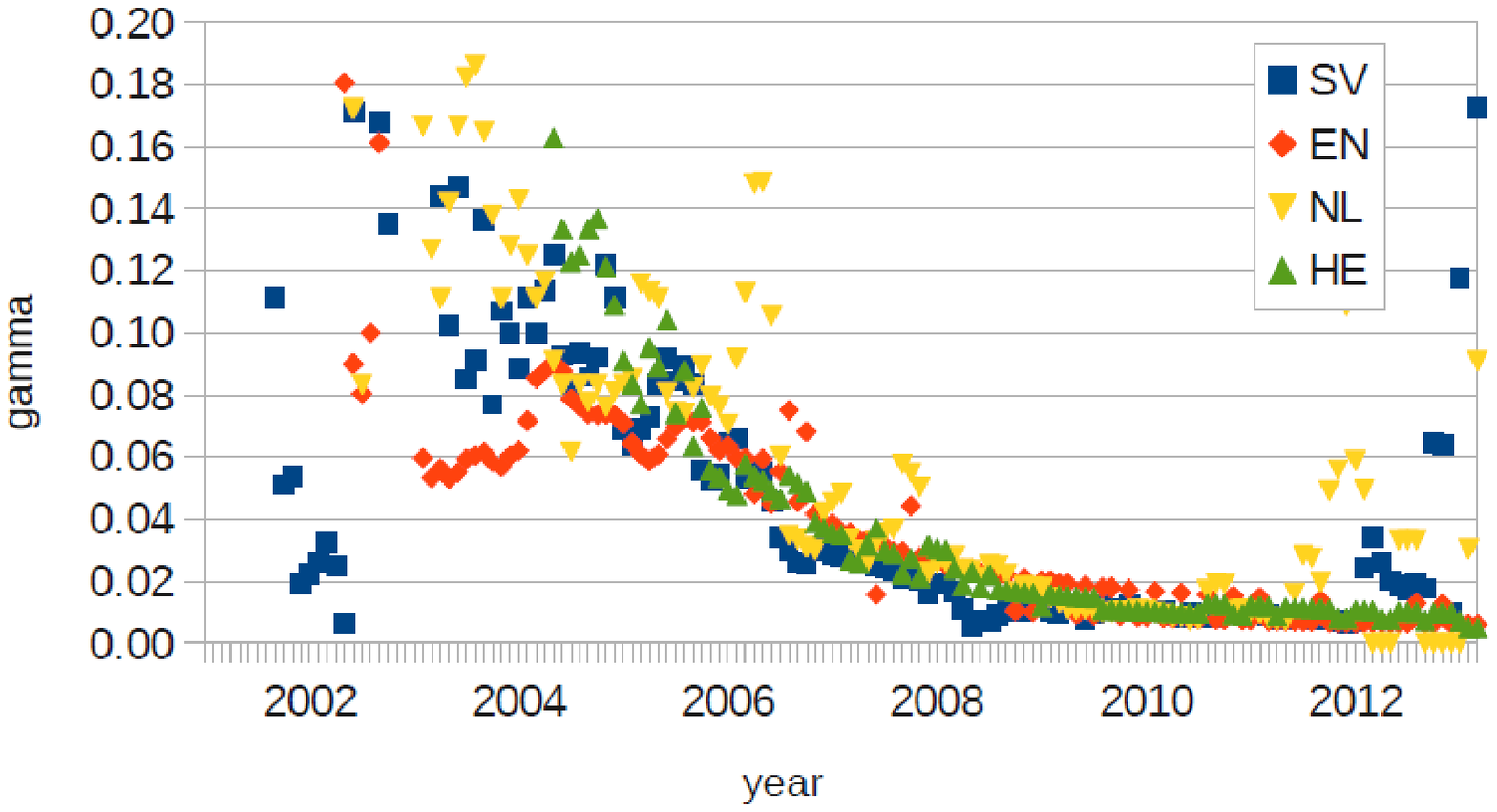}
\caption{For the Wikipedia projects in Swedish (blue), English (red), Dutch (yellow), and 
Hebrew (green) (a) the number of pages and (b) the exponential growth rate $\gamma$ is shown.}
\end{figure}

Figure 16(a) shows the total number of pages for four Wikipedia projects (Swedish, English, 
Dutch, and Hebrew).  The number of pages $N_P(t)$ is growing by the number of new pages 
$n_P(t) = N_P(t) - N_P(t-1)$ per time interval $\Delta t = 1$ month.  Figure 16(b) shows the 
growth rate $\gamma$ for an exponential growth model $N_P(t) = N_P(t-1) \exp(\gamma)$, 
which has been determined by $\gamma \approx n_P(t)/N_P(t)$.  Note that an increased $\Delta t$
has been used if $n_P(t)=0$.

In the beginning the growth rate $\gamma$ is quite large.  Later, a tendency towards saturation 
can be identified.  This shows that the character of edit events changed over time.  In the 
early stage of a Wikipedia project most of the edit events are related to the creation of new 
pages, while later on the internal structure evolves.  For the English Wikipedia project, one 
can see an intermediate regime with an exponential growth ($\gamma \approx 0.7$).  Such an 
exponential growth cannot be unambiguously identified for the Swedish Wikipedia.  Interestingly,
the page-growth rate has been drastically increasing during the last few months (in 2013) for 
the Dutch and -- even more dramatically -- for the Swedish Wikipedia.  Acutally, the Swedish 
and the Dutch Wikipedia started to create articles using bots. 

\begin{figure}[t]
\includegraphics[width=0.9\hsize]{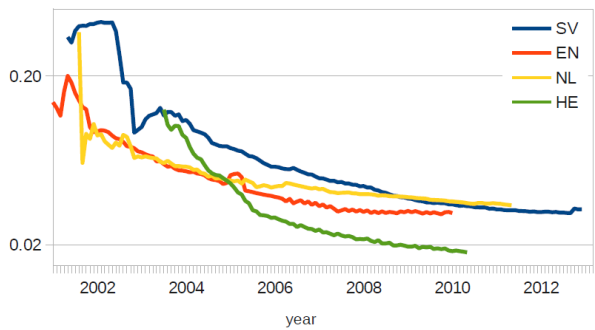}
\caption{Ratio of total number of pages $N_P(t)$ and total number of links $l_P(t)=l_{\rm 
int}(t) + l_{\rm ext}(t)$ (internal and external) as fuction of time between January 2001 and 
April 2013 for the Wikipedia projects in Swedish (blue), English (red), Dutch (yellow), and 
Hebrew (green).  Note that only Swedish data is available till present and that the vertical
axis has a logarithmic scale.}
\end{figure}

Figure 17 confirms that editorial activity tends to focus more on the addition of links than
the creation of new pages during later states of Wikipedia evolution.  It shows the ratio of 
the total number of pages $N_P(t)$ and the total number of links $l(t)$ as function of time. 
For all languages this ratio decreases during most of the time after a relatively large value
(around 0.2, i.e., approximately five links per article) in the beginning.  The final values
are between 0.015 and 0.04, i.e. at approximately 25-60 links per article.  For the Dutch and 
the Swedish Wikipedia the initial change (between 2001 and 2003) is quite sudden.  In general,
all four languages show a stronger decay of the page number to link number ratio in the beginning 
and a much slower decay later on.  This behavior suggests that an exponential decay model may 
also be appropriate.  However, we cannot find any regimes with unique or approximately constant 
decay rates for any of the considered four languages.  The different decay rates of the page 
number to link number ratio might also be indicators for two different network growth processes.

The Swedish Wikipedia has initially $\approx 5$ links per page and later the number of links 
per page increases to an average of $\approx 25$.  This is in line with the change in the 
degree distribution, which is shown in Fig. 15.  Here one can see a continuous shift 
towards a dominating structural growth process, while the growth of content volume -- measured 
in number of pages -- becomes less important.  The current ratio of page number to link number
for the Swedish Wikipedia is quite similar those for the English and the Dutch version, while 
the Hebrew Wikipedia has about twice as many links per article.  During the quick growth of the 
Swedish Wikipedia article number in the past few months (see Figs.~16(a,b)), the article to link 
number ratio has slightly grown, which may indicate a slight change of the structure towards 
properties typical for Wikipedias at earlier stages of evolution.  Although, this weak growth 
is still comparable with typical fluctuations of the ratio (just about twice as large), it may
indicate that creating articles by bots leads to a step back in the quality of content.

\begin{figure}[t]
\includegraphics[width=0.9\hsize]{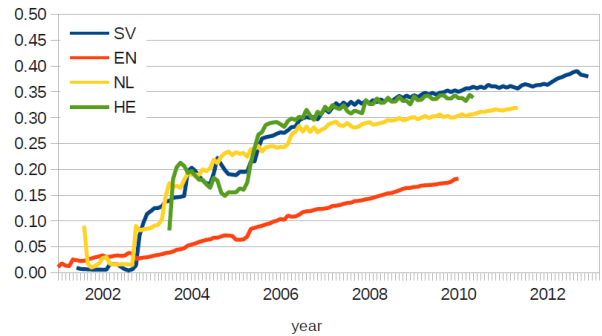}
\caption{Ratio of number of external links $l_{\rm ext}(t)$ and total number of links $l_P(t)=
l_{\rm int}(t) + l_{\rm ext}(t)$ (internal and external) as fuction of time between January 
2001 and April 2013 for the Wikipedia projects in Swedish (blue), English (red), Dutch (yellow), 
and Hebrew (green).  Note that only Swedish data is available till present.}
\end{figure}

Next we separate the changes of internal and external link numbers.  External links (to other
language versions or references outside Wikipedia) are particularly important for confirmation 
of the article content and can thus be regarded as an important quality indicator for the 
articles.  Figure 18 shows the ratio of the number of external links to the number of all
links (internal and external).  The increasing curves show, that the ratio of external 
links grows for most of the time in all four Wikipedias.  We note that there are two major 
groups of external links: just 'further reading' links (often in bad articles) and references 
(more likely in good articles).  The habit of adding references increased in the last years, 
while one got more discouraged adding just simple links; they are usually also limited to 
3-5 per article.

In Fig. 18 one can also find 
indicators for two regimes for each Wikipedia project except the English: fast growth from 
approximately 2003 to 2005 followed by a behavior close to saturation after 2005.  Table 2 
shows the times where the qualitative behavior changes.  The transition time A ($t_A$) is 
determined from Figure 17, which shows the ratio of total number of pages and total number 
of links while the transition time C ($t_C$) is based on the plot in Figure 18, which shows 
the ratio of external and internal links.  For all languages we find that $t_A$ is before 
$t_C$, but the differences vary from 4 to 52 months depending on the languages.  
 
\begin{center}
\begin{tabular}{c|c|c|c}
Language & $t_A$ & $t_C$ & $t_C - t_A$\\ \hline
SV & 05/2003 & 10/2007 & 43\\
EN & 12/2001 & 12/2002 & 12\\
NL & 01/2002 & 05/2006 & 52\\
HE & 09/2005 & 01/2006 & 4
\end{tabular}
\label{Tab:2}
\end{center}

The Swedish Wikipedia has already reached the largest ratio of external links among all links 
in 2010 and has continued to increase this ratio during the last 2.5 years (see Fig. 18).  
This is an indication for a very good reference quality of average articles in the Swedish 
Wikipedia, better than in the Dutch Wikipedia.  Note that the ratio of external links is very 
much lower in the English Wikipedia, just approximately half as large as in the Swedish 
Wikipedia (data from 2010).  The slight drop of the Swedish curve in Fig. 18 in the last 
months is probably associated with the drastic increase of the total number of articles 
(see Fig. 16).  However, it is too weak to be considered as an indication of a drop in 
article reference quality, and there was a significant larger increase during 2012 just 
before the slight drop.  We note that  bot generated articles usually have a quite high 
density of references, meaning just one sentence but 2-3 references to publications which, 
however, may not be linked using a web link.  Overall, we can conclude that the article 
reference quality of the Swedish Wikipedia has not decreased significantly in the past months 
despite the quite drastic content increase with nearly doubled article number probably due to 
article creation by bots.

\section{Literature overview} \label{sec:Literature review}

The following list has been compiled in December 2012.

\begin{enumerate}
\item \underline{In Wikipedia:} \textbf{stats.wikimedia.org}
	\begin{itemize}
	\item[$\bullet$] Monthly: 
	\subitem - Unique Visitors per Region
	\subitem - Wikimedia Projects Reach by Region
	\item[$\bullet$] Special:
	\subitem - Growth per Major Project
	\subsubitem \textit{Project = Wikibooks, Wikinews, Wikipedia, Wikiquote, Wikisource, 
        Wikiversity, Wikispecial, Wiktionary}
	\subitem - Growth per Language Edition
	\subitem - Summaries for all Wikimedia Wikis
	\subitem - Page Views per Project
	\subitem - Wikipedia Page Views / Page Edits per Country / Language
	\subitem - Breakdown of Page Views for all Wikimedia Projects
	\subitem - Server Requests
	\subitem - Page Edits \& Reverts
	\subitem - User Feedback (with Moodbars)
	\subitem - Current Status
	\subitem - Overview Recent Month
	\subitem - \dots
	\item[$\bullet$]External:
	\subitem - Visualizations, an overview
	\subitem - Google Trends
	\subitem - Page Views-Trending Topics
	\subitem - Page Views \textbf{stats.grok.se / Emw}
	\subitem - Most Edited Pages-Wikirage
	\subitem - Log Analysis Oct 2007-Dragons flight
	\subitem - \dots
	\end{itemize}
\item \underline{List of publications} \\ \textbf{\tiny 
http://de.wikipedia.org/wiki/Wikipedia:Wikipedistik/Arbeiten}
	\begin{itemize}
	\item[$\bullet$] Opuszko, Marek et al. (2010): Qualit\"atsmessung in der Wikipedia. 
Ein Ansatz auf Basis von Markov-Modellen. In: Schumann, Matthias et al. (Hrsg.): 
Multikonferenz Wirtschaftsinformatik 2010. G\"ottingen: Universit\"atsverlag G\"ottingen, 
S. 705-716.
	\item[$\bullet$] Projekt Missbrauch Sozialer Software im Allgemeinen, automatische 
Erkennung von Wikipedia-Vandalismus und Edit-Wars im Speziellen. WWW: Bauhaus Universit\"at 
Weimar, Web Technology \& Information Systems
	\item[$\bullet$] Graham, Marc (2011): Mapping Wikipedia's augmentations of our planet. 
Evaluation of the articles' Geotags in seven language versions with respect to their world-wide
distribution.
%Auswertung der Artikel-Geotags in sieben verschiedenen Sprachversionen nach ihrer weltweiten 
%Verteilung.
	\item[$\bullet$] Marcel Minke studies in his doctoral thesis if the sequence of Wikipedia
	articles viewed by users can lead to conclusions for their areas of knowledge (Institut 
f\"ur Mathematik und Angewandte Informatik, Universit\"at Hildesheim, Germany).
%ob sich bei einem 
%teilweise vorgegebenen Navigationspfad anhand der Navigation der Benutzer innerhalb der 
%Wikipedia R\"uckschl\"usse auf deren Wissensdom\"anen ziehen lassen.
	\end{itemize}
\item \underline{Further articles} 
\begin{itemize}
	\item[$\bullet$] K. Stein and C. Hess, Viele Autoren, gute Autoren? Eine Untersuchung 
ausgezeichneter Artikel in der deutschen Wikipedia, in: Web 2.0 — Eine empirische Bestandsaufnahme, 
editors: P. Alpar and S. Blaschke, pages 105-129.
{\tiny http://www.springerlink.com/content/978-3-8348-0450-1/\#section=214918\&page=5\&locus=0}
	\item[$\bullet$] D. Kinzler, Assessing the quality of Wikipedia articles with lifecycle 
based metrics, conference paper.
	\item[$\bullet$] A. Dalby, Wikipedia(s) on the language map of the world, English Today 23, 
3 (2007)\\ {\tiny http://journals.cambridge.org/action/displayAbstract?fromPage=
online\&aid=1036104}
    \item[$\bullet$] U. Pfeil, P. Zaphiris, C. S. Ang, Cultural differences in collaborative 
 authoring of Wikipedia, J. Computer-Mediated Communication 12, 88 (2006)\\
{\tiny http://onlinelibrary.wiley.com/doi/10.1111/j.1083-6101.2006.00316.x/full}
	\end{itemize}
\end{enumerate}

\end{document}